%% file: art_ch3csnh2_astro_ab.tex
\documentclass{aa}  

\usepackage{graphicx}
\usepackage{txfonts}
\usepackage[version=4]{mhchem}
\usepackage{natbib}

\usepackage{booktabs}
\usepackage{amsmath}
\usepackage{enumitem}

\bibpunct{(}{)}{;}{a}{}{,} 

\begin{document}

   \title{Expanding the submillimeter wave spectroscopy and astronomical search for thioacetamide (\ce{CH3CSNH2}) in the ISM}


\author{A. Remijan \inst{1}, C. Xue \inst{2,1}, L. Margul\`es \inst{3}, A. Belloche \inst{4}, R. A. Motiyenko \inst{3}, J. Carder\inst{2},  C. Codella \inst{5}, N. Balucani \inst{6,5,7}, C. L. Brogan \inst{1}, C. Ceccarelli \inst{6}, T. R. Hunter \inst{1}, A. Maris \inst{8}, S. Melandri \inst{8}, M. Siebert, \inst{9,1} B. A. McGuire \inst{10,1}}

    \institute{
         National Radio Astronomy Observatory, Charlottesville, VA 22903, USA\\
         \email{aremijan@nrao.edu}
         \and Department of Chemistry, University of Virginia, Charlottesville, VA 22903, USA
         \and Univ. Lille, CNRS, UMR 8523 - PhLAM - Physique des Lasers Atomes et Mol\'ecules, 59000 Lille, France
         \and Max-Planck-Institut f\"{u}r Radioastronomie, Auf dem H\"{u}gel 69, 53121 Bonn, Germany
         \and INAF, Osservatorio Astrofisico di Arcetri, Largo E. Fermi 5, 50125 Firenze, Italy
         \and Universit\'e Grenoble Alpes, CNRS, Institut de Plan\'etologie et d’Astrophysique de Grenoble(IPAG), F-38000 Grenoble, France
         \and Dipartimento di Chimica, Biologia e Biotecnologie, Universit\'a di Perugia, Via Elce di Sotto 8, I-06123 Perugia, Italy
         \and Dipartimento di Chimica “Giacomo Ciamician”, Universit\'a di Bologna,, Via Selmi 2, I-40126, Bologna, Italy
         \and Department of Astronomy, University of Virginia, Charlottesville, VA 22903, USA
         \and Department of Chemistry, Massachusetts Institute of Technology, Cambridge, MA 02139, USA}

   \date{Received XXXXX XX, 2020; accepted XXXXX XX, XXXX}

 
  \abstract
   {One of the biggest unsolved mysteries of modern astrochemistry is understanding chemical formation pathways in the interstellar medium (ISM) and circumstellar environments (CSEs).  The detections (or even nondetections) of molecules composed of low-abundance atomic species (such as S, P, Si, and Mg) may help to constrain chemical pathways. Thioacetamide (\ce{CH3CSNH2}) is the sulfur analog to acetemide (\ce{CH3CONH2}) and it is a viable candidate to search for in astronomical environments - specifically toward regions where other S-bearing molecules have been found and, if possible, that also contain a detection of \ce{CH3CONH2}.
If detected, it would not only continue to expand the view of molecular complexity in astronomical environments, but also help to better elucidate the possible formation pathways of these types of species in these environments.}
   {Our aim is to expand the frequency range of the measured rotational spectrum of \ce{CH3CSNH2} beyond  150 GHz and then to use those measurements to extend the search for this species in the ISM. 
   The new laboratory measurements and expanded search cover more parameter space for determining under what conditions \ce{CH3CSNH2} may be detected, leading to possible constraints on the formation of large S-bearing molecules found in the ISM.}
   {The rotational spectrum of \ce{CH3CSNH2} was investigated up to 650 GHz. Using the newly refined spectrum of \ce{CH3CSNH2}, as well as additional spectroscopic data on the chemically related species \ce{CH3CONH2}, a variety of astronomical sources were searched including data from the following large surveys: Prebiotic Interstellar Molecule Survey (PRIMOS) conducted with the Green Bank Telescope (GBT); Exploring molecular complexity with ALMA (EMoCA) conducted with the Atacama Large Millimeter/submillimeter Array (ALMA); and Astrochemical Surveys At IRAM (ASAI) conducted with the Institut de Radioastronomie Millim\'etrique (IRAM) 30m Telescope.}
   {A total of 1428 transitions from the v$_t$=0 state with maximum values J=47 and K$_a$=20  in  the  range up  to  330  GHz,  and J=95 and K$_a$=20  in  the  range from  400–660 GHz were assigned. We also assigned 321 transitions from the v$_t$=1 state with the maximum values J=35 and K$_a$=9 up to 330 GHz. We achieved a final fit with a root-mean-square deviation of 43.4 kHz that contains 2035 measured lines from our study and the literature for v$_t$=0 and v$_t$=1 states of A and E symmetries. 
   The final fit is based on  the rho-axis-method (RAM)  Hamiltonian  model  that  includes  40  parameters.  An astronomical search for \ce{CH3CSNH2} was conducted based on all the new spectroscopic data.  No transitions of \ce{CH3CSNH2} were detected toward any of the sources contained in our survey.  Using the appropriate telescope and physical parameters for each astronomical source, upper limits to the column densities were found for \ce{CH3CSNH2} toward each source.}
   {}

   \titlerunning{Rotational spectrum and ISM search of Thioacetamide}
 \authorrunning{Remijan et al.}

   \maketitle
%

\section{Introduction}

One of the biggest unsolved mysteries of modern astrochemistry is understanding chemical formation pathways in the interstellar medium (ISM) and circumstellar environments (CSEs). The detections (or even nondetections) of new species are necessary to help calibrate and constrain existing chemical models in an attempt to make them more predictive. Using low-abundance atomic species (such as S, P, Si, and Mg) may help to constrain the chemical pathways.  This is because routes to forming molecules from these elements are typically more limited than from species such as C, O, N, and H that 1) form molecules easily; 2) are in large abundance; and 3) show a rich molecular chemistry. Furthermore, while molecules that formed from these low-abundance atomic species have primary formation pathways in the gas phase, for example, \ce{SiO} \citep{2017JPCA..121.1675R}, \ce{SiS} \citep{2021arXiv210903367M}, and \ce{HF} \citep{2009ApJ...706.1594N}, grain surface chemistry is critical because grain surfaces provide sites of absorption and desorption where atomic species can interact and dissipate excess energy as a result of their formation.  In addition, molecular species containing sulfur have been postulated to be used as ``chemical clocks" to measure the overall evolutionary state of an interstellar environment \citep[for example,][]{2001ASSL..253.....M, 2004AnA...422..159W, 2015ApJ...802...40L}. Sulfur is also extensively used in observations of solar system bodies such as Mars, Venus, Io, and meteorites to elucidate chemical and physical processes and the effects of solar radiation on these objects \citep{1999AJ....118.1850B, 2010Icar..208..353M, 2012Icar..217..740B, FRANZ2019119}. Sulfur bearing species are also sensitive to gas temperature and density and, as such, have been used to understand the physical environments of numerous astronomical environments \citep[][and references therein]{1997ApJ...481..396C,2014A&A...567A..95E, 2018A&A...617A..28S}.

\begin{table}[h!]
\centering
\label{Sspecies}
\caption{Current list of known ISM/CSE S-bearing species.  Entries in {\bf bold} currently do not have a detected O-substituted counterpart in the ISM/CSE.}
\begin{tabular*}{\columnwidth}{@{\extracolsep{\fill}}llllll}
\hline \hline
2- & 3-  & 4-  & 5-  & 6-  & 9-atoms\\ 
\hline
\ce{CS & C2S & C3S  & HC3S+ & CH3SH & CH3CH2SH} \\
\ce{NS & H2S & H2CS & H2C2S} & {\bf \ce{C5S}}  &\\
\ce{SH & HCS & HNCS} & {\bf \ce{C4S}}   & {\bf \ce{H2C3S}}&\\
\ce{SO & HSC & HSCN & HC(O)SH & HCSCCH} & \\
\ce{SiS& OCS & HCCS & HCSCN} & &\\
\ce{NS+& SO2} & & & &\\
\ce{SH+& S2H} & & & &\\
\ce{SO+& HCS+}& & & &\\
& \ce{NCS}& & & & \\
\hline
\end{tabular*}
\end{table}

However, the main sulfur carriers on grain mantles have yet to be fully characterized. Only OCS and \ce{SO2} have been unambiguously identified on grain surfaces \citep{1995ApJ...449..674P, 1997ApJ...479..839P, 2009ApJ...694..459Z}. Hydrogen sulfide (\ce{H2S}) - a main reactant in numerous chemical networks on grain surfaces - has yet to be identified \citep{2003A&A...412..133V}.

From the list of known interstellar S-bearing molecules (Table~\ref{Sspecies}), nearly all species have corresponding detections of their O-substituted counterparts. For example, methanol (\ce{CH3OH}) is the O-analog to methyl mercaptan (\ce{CH3SH}). In fact, the only analogous O-bearing molecules that have yet to be detected from the list are diacetylene ether (\ce{C4O}), diacetylene ketone (\ce{C5O}) and propadienone (\ce{CH2CCO}). In addition, from a formation chemistry and detection point of view, it is very surprising that (\ce{CH2CCS}) \citep{2021A&A...648L...3C} has been detected as well as its thioaldehyde isomer - \ce{HCCCHS} \citep{2020A&A...642A.206M}.  This is completely opposite from what is found in the O-analog species.  That is, the aldehyde \ce{HCCCHO} is very abundant and easily detected toward several environments \citep{1988ApJ...335L..89I, 2004ApJ...610L..21H, 2008ApJ...672..352R} whereas \ce{CH2CCO} has so far eluded detection, though recent theoretical predictions offer a possible explanation \citep{2019ApJ...878...80S}.  As such, searching for S-bearing species with already identified O-substituted counterparts in astronomical environments can provide a basis (or proxy) for searching for more complex species.  And, while it may not always lead to a detection -- or sometimes a completely different species may be found -- searching for and setting detection limits will help to calibrate existing chemical networks that predict abundances of S-bearing species. 

Acetamide (\ce{CH3CONH2}) and N-methylformamide (\ce{CH3NHCHO}), are two isomers which contain a peptide bond \citep{2006ApJ...643L..25H, 2017A&A...601A..49B, 2020ApJ...901...37L}. The detection of these molecules under interstellar conditions is important because 1) it again increases the limit of molecular complexity found in astronomical environments and 2) these molecules may provide a starting point to the formation of larger, more complex, prebiotic astronomical molecules. And, to date, the formation routes to these species are still uncertain. Thioacetamide (\ce{CH3CSNH2}) is the sulfur analog to \ce{CH3CONH2} and is a viable candidate to search for in astronomical environments - specifically toward regions where other S-bearing molecules have been found and, if possible, that also contain a detection of \ce{CH3CONH2}. If detected, it would not only continue to expand the limits of molecular complexity in astronomical environments but also help to better elucidate the possible formation pathway of these types of species in astronomical environments. 

Thioacetamide is a planar, asymmetric top molecule with C$_s$ symmetry. Its rotational spectrum was recorded recently by free-jet absorption spectroscopy in the 59.6–110.0 GHz frequency region \citep{2019ECS.....3.1537M}. In addition to detecting the most abundant isotopologue, the \ce{^{34}S} species was also observed in natural abundance ($4.25\%$). Some of the rotational transition lines showed small splittings (1-2 MHz) due to the \ce{^{14}N} nuclear quadrupole interaction. Because of the internal rotation of the methyl group, all lines are further split into A and E components.  The barrier to this internal rotation is low ($V_3 \approx  110\,\mathrm{cm}^{-1}$), which leads to the relatively large frequency splittings between the A and E states. That study also conducted the first attempt at an astronomical search for this molecule using the observational data from the unbiased Astronomical Surveys at IRAM (ASAI\footnote{\url{http://www.oan.es/asai/}}) program \citep{2018MNRAS.477.4792L} toward two sources (associated with low-mass star forming regions) that illustrated diverse and extreme physical conditions - L1544, a cold prestellar core, and L1157-B1, a shocked region within the ISM. These sources were also selected because the emission regions are known to be extended with respect to the IRAM 30m beam (10$\arcsec$--30$\arcsec$), thus minimizing the affects of beam dilution. The observations were carried out at 3 mm (72–116 GHz), 2 mm (126–173 GHz), and 1.3 mm (200–276 GHz). 

No detection of \ce{CH3CSNH2} was reported and upper limits were given on the abundance of \ce{CH3CSNH2} toward these sources. And while several smaller sulfur bearing species have been detected toward these regions, no larger S-bearing molecules have been detected and there has also not been a detection of \ce{CH3CONH2} - the O-analog of \ce{CH3CSNH2} toward these sources either.

In order to conduct a broader radio astronomical search for \ce{CH3CSNH2}, additional spectroscopic data (both from the laboratory and space) from a wider range of sources are necessary. As such, we present here a new study investigating the rotational spectrum of \ce{CH3CSNH2} using laboratory measurements from 150 to 650 GHz. These measurements were done at room temperature which enabled a global analysis of both the ground and the first torsional excited states. This new study has permitted, for the first time, an accurate prediction in the millimeter- and submillimeter wave ranges necessary for a more extensive astronomical search.  In addition, the existing lower frequency data were used to search for transitions at cm wavelengths that may potentially show maser activity. In Section~\ref{Sec:Spectroscopy} we describe the experimental details of the study and the full analysis of the spectrum; in Section~\ref{Sec:Observation} we describe the astronomical search for this species specifically toward regions with previous detections of \ce{CH3CONH2} as well as other sources obtained from archival data from various astronomical facilities; in Section~\ref{Sec:Discussion} we discuss the implication of the investigation in the context of the current presumed formation pathways of \ce{CH3CONH2} in the ISM and a comparison of the abundances found between the S- and O-substituted species already detected toward various astronomical environments and finally, in Section~\ref{Sec:Conclusion} we provide a summary and possible future observational work needed to better elucidate the formation of large S-bearing species in the ISM.

\section{Spectroscopic study \label{Sec:Spectroscopy}}

\subsection{Experimental details \label{Sec:experimental details}}

The sample of \ce{CH3CSNH2} was purchased from Aldrich and used without purification. The absorption measurements of the rotational spectrum between 150 and 650 GHz were conducted using the fast-scanning Terahertz spectrometer in Lille. The details of the spectrometer are described in \citet{2015JMoSp.317...41Z}.

The radiation source present in the spectrometer is a commercially available Virginia Diodes (VDI) frequency multiplication chain driven by an in-house fabricated fast sweep frequency synthesizer. The fast sweep system is based on the up-conversion of an AD9915 direct digital synthesizer (DDS) operating between 320 and 420 MHz into the Ku band (12.5–18.25 GHz) by mixing the signals from the AD9915 and an Agilent E8257 synthesizer with subsequent sideband filtering. The DDS provides rapid frequency scanning with up to 50 ms per point frequency switching rate. In order to obtain an optimum signal-to-noise ratio, the spectrum was scanned with a slower rate of 1 ms per point and with 4 scans co-averaged. The sample pressure during measurements was $\sim15$ Pa at room temperature. Absorption signals were detected by an InSb liquid He-cooled bolometer (QMC Instruments Ltd.). Estimated uncertainties for the measured line frequencies are 30 kHz.

\subsection{Analysis of the spectra \label{Sec:analysis of the spectra}}

Thioacetamide is an asymmetric near-prolate rotor ($\kappa \approx$ -0.48). It has nonzero dipole components along the $a$- ($\mu_a$ = 4.1~D) and $b$-axes ($\mu_b$ = -1.1~D) as determined by \citet{Sundaresa1957}. The molecule is challenging from the spectroscopic point of view as it exhibits large amplitude methyl torsional motion. The interaction between internal rotation of the methyl group and overall rotation of the molecule complicates the spectral analysis. For \ce{CH3CSNH2}, the barrier to internal rotation is low ($V_3 \approx 110~\mathrm{cm}^{-1}$), and leads to relatively large torsional $A-E$ splittings of the rotational energy levels. The low-lying excited torsional states add more complexity through Coriolis-type interactions with the ground state as well as with each other. As a result, a correct interpretation of the ground state rotational spectrum of \ce{CH3CSNH2} within a large range of $J$ and $K_{a}$ quantum numbers requires a global analysis including excited torsional states.

In the present study, we used the theoretical model base of the rho-axis method (RAM) \citep{1962JChPh..37.2516K,1968JChPh..48.5299L,1984JMoSp.108...42H}. The main advantage of the RAM Hamiltonian is its general approach that simultaneously takes into account the $A$- and $E$-symmetry species and all torsional levels, intrinsically taking the inter-torsional interactions into account within the rotation-torsion manifold of energy levels. This method is particularly suitable for a difficult case of low barrier to internal rotation \citep{ILYUSHIN2004, 2012ApJ...755..153N} or when large coupling between internal and overall rotation occurs \citep{2014JMoSp.295...44S}, or both. Because this method has been presented in great detail in the literature \citep{1994JMoSp.163..559H, 2010JMoSp.260....1K}, we do not describe it here. The RAM used for the spectral analysis in this present study has been previously shown to be very effective for a number of molecules that contain an internal methyl rotor \citep{2010JMoSp.259...26I,2013JMoSp.290...31I,2014JMoSp.295...44S}. In particular, it has been used for the oxygen-bearing analog of \ce{CH3CSNH2}, \ce{CH3CONH2} \citep{2004JMoSp.227..115I} which has a lower barrier to internal rotation than \ce{CH3CSNH2} ($V_3 \approx 25~\mathrm{cm}^{-1}$). For the analysis, we employed the RAM36 (rho-axis-method for 3- and 6-fold barriers) code that uses the RAM approach for molecules with a C$_{3v}$ top attached to a C$_s$ or C$_{2v}$ symmetry molecular frame and having 3- or 6-fold barriers to internal rotation, respectively. A general expression for the RAM Hamiltonian implemented in this code, as well as further details of the theoretical approach, is found in \citet{2010JMoSp.259...26I} and \citet{2013JMoSp.289...41I}. The code RAM36 used for this analysis is publicly available\footnote{\url{http://www.ifpan.edu.pl/~kisiel/introt/introt.htm}}, and has been successfully utilized in the spectral analysis of many molecules \citep{2010JMoSp.259...26I,2013JMoSp.290...31I,2014JMoSp.295...44S}, including the recent analysis of the sulfur-containing molecule \ce{CH3CHS} \citep{2020JMoSp.37111304M}.

To begin, we refit the data from the previous study \citep{2019ECS.....3.1537M} using the RAM36 program. Using predictions obtained from the initial RAM36 fit, we assigned and fit the \textbf{$^aR$} transitions in the usual manner, with numerous cycles of refinement of the parameter set while new data were gradually added. However, it was not possible to observe the hyperfine splittings due to our Doppler-limited resolution. 

Next, we searched for the $^bR$ and $^bQ$ lines. However, these transitions could not be assigned unambiguously due to the dense spectra and their much smaller relative intensities. Also, several high K$_a$ series of the ground torsional state transitions were perturbed by the interactions with the first excited torsional state; therefore it was necessary to get more precise information about the positions of energy levels of this state. The band origins of the first torsional states are $\nu^A = 61.6589~\mathrm{cm} ^{-1}$, $\nu^E = 51.3985~\mathrm{cm}^{-1}$, giving relative intensities to ground state at room temperature of 0.74 and 0.78, respectively. The analysis of the first torsional state also stabilized the fit and reduced the correlation between torsional parameters of the RAM Hamiltonian models. 

Finally, we assigned 1428 transitions from the $v_t = 0$ state with the maximum values $J = 47$ and $K_a = 20$ in the range up to 330~GHz, and $J = 95$ and $K_a = 20$ in the range 400--660~GHz. We also assigned 321 transitions from the $v_t = 1$ state with the maximum values $J=35$ and $K_a = 9$ up to 330~GHz. The final data set contains both the data from the previous study \citep{2019ECS.....3.1537M}, and new assignments in the millimeter- and submillimeter wave ranges. The total of 2035 lines for $v_t =0$ and $v_t =1$ states of $A$ and $E$ symmetries were fitted with root-mean-square deviation of 43.4~kHz (wrms=0.90). The final fit is based on the RAM Hamiltonian model that includes 40 parameters. The values of the molecular parameters obtained from the final fits are presented in Table~\ref{tabpar}, where they are compared with the parameters from the initial fit of the data from the previous study \citep{2019ECS.....3.1537M}. The new combination of data from the $v_t = 0$ and $v_t =1$ states required 23 more parameters that also improved the overall quality of the fit. In particular, the inclusion of the $v_{t}=1$ state permitted the removal of the strong correlation between $F$ and $\rho$ parameters of the Hamiltonian, and to accurately determine the V$_6$ parameter.

\begin{table*}
\caption{Molecular parameters in $\mathrm{cm}^{-1}$ of \ce{CH3CSNH2} obtained with the RAM36 program.}
\centering  
\label{tabpar}           
\centering   
\begin{tabular*}{\textwidth}{@{\extracolsep{\fill}}ccccc}
\hline\hline
ntr$^{\mathrm{a}}$ & Parameter$^{\mathrm{b}}$ & Operator$^{\mathrm{c}}$ & Value$^{\mathrm{d}}$ & Value (this study)\\ 
\hline
220 & $F    $              &  $p_\alpha^2       $                                               &   5.650542416$^{\mathrm{e}}$                 &   5.45339(10)$^{\mathrm{f}}$                              \\
220 & $V_3  $              &  $\frac{1}{2}(1-\cos3\alpha)$                                      & 108.860(66)                        & 110.71340(94)                            \\
211 & $\rho $              &  $J_zp_\alpha      $                                               &  0.042368(54)                       &  0.0437894(17)                           \\
202 & $A_{RAM}$            & $ J_z^2 $                                                          &  0.270791(38)                       &  0.2696379(20)                           \\
202 & $B_{RAM}$            &  $J_x^2                                          $                 &  0.2207334(66)                      &  0.22085611(92)                          \\
202 & $C_{RAM}$           &  $J_y^2                                        $                   &  0.11213403(48)                     &  0.112127546(92)                         \\
202 & $D_{zx}     $        &  $\lbrace J_z,J_x \rbrace                      $                   &    $-$0.0749884(24)                 &    $-$0.07505330(80)                      \\
440 & $V_6        $        &  $\frac{1}{2}(1-\cos 6\alpha)                  $                   &  $                                 $     &   $-$14.66262(38)                         \\
422 & $F_J        $        &  $J^2p_\alpha^2                                $                   &  $                                 $      &  $ -0.456(12)\times 10^{-6}   $          \\
422 & $F_K        $        &  $J_z^2p_\alpha^2                              $                   &  $                                 $     &  $  -0.484(13)\times 10^{-5}   $        \\
422 & $F_{zx}     $        &  $\frac{1}{2}p_\alpha^2\lbrace J_z,J_x \rbrace $                   &  $                                $     &  $  -0.507(93)\times 10^{-6}   $        \\
422 & $V_{3J}     $        &  $J^2(1-\cos 3\alpha)                          $                   &  $ 0.2781(80)\times 10^{-3}   $     &  $ 0.14296(43)\times 10^{-3}   $        \\
422 & $V_{3K}     $        &  $J_z^2(1-\cos 3\alpha)                        $                   &  $ 0.149(12)\times 10^{-2}   $     &  $-0.14891(40)\times 10^{-2}   $        \\
422 & $V_{3zx}    $        &  $\frac{1}{2}(1-\cos 3\alpha)\lbrace J_z,J_x \rbrace $             &  $ 0.1820(14)\times 10^{-2}   $     &  $ 0.20727(28)\times 10^{-2}   $       \\
422 & $V_{3xy}    $        &  $(J_x^2-J_y^2)(1-\cos 3\alpha)                  $                 &    $0.2311(85)\times 10^{-3}   $       &   $ 0.9789(38)\times 10^{-4}   $         \\
422 & $D_{3zy}    $        &  $\frac{1}{2}\sin 3\alpha\lbrace J_z,J_y \rbrace  $                &  $                           $     &  $  0.9823(63)\times 10^{-3}   $       \\
422 & $D_{3xy}    $        &  $\frac{1}{2}\sin 3\alpha\lbrace J_x,J_y \rbrace  $                &  $                           $       &  $ 0.450(32)\times 10^{-4}   $         \\
413 & $\rho_J     $        &  $J^2J_zp_\alpha                               $                   &  $    -0.229(15)\times 10^{-5}  $    &  $  -0.3828(59)\times 10^{-6}   $      \\
413 & $\rho_K     $        &  $J_z^3p_\alpha                                $                   &  $    0.1761(49)\times 10^{-4}   $     &  $  0.2695(19)\times 10^{-5}   $       \\
413 & $\rho_{zx}  $        &  $\frac{1}{2}p_\alpha \lbrace J_z^2,J_x \rbrace $                  &  $                              $     &  $   -0.982(24)\times 10^{-6}  $       \\
404 & $D_{zxJ}      $        &  $\frac{1}{2}\lbrace J_z,J_x \rbrace J^2                  $                    &  $                             $      &  $ -0.433(44)\times 10^{-7}   $       \\
404 & $D_{zxK}      $        &  $\frac{1}{2}\lbrace J_z,J_x \rbrace J_z^2                $                    &  $                             $      &  $ 0.6597(45)\times 10^{-6}   $       \\
404 & $\Delta_J   $        &  $-J^4                                         $                    &  $                             $      &  $ 0.1137(12)\times 10^{-6}   $       \\
404 & $\Delta_{JK}$        &  $-J^2J_z^2                                    $                    &  $                              $     &  $ -0.3446(24)\times 10^{-6}   $       \\
404 & $\Delta_K   $        &  $-J_z^4                                       $                    &  $                              $     &  $  0.6353(49)\times 10^{-6}   $       \\
404 & $\delta_J   $        &  $-2J^2(J_x^2-J_y^2)                           $                    &  $                              $     &  $  0.1024(12)\times 10^{-6}   $      \\
404 & $\delta_K   $        &  $-\lbrace J_z^2,(J_x^2-J_y^2) \rbrace         $                    &  $                             $      &  $ 0.6668(99)\times 10^{-7}   $       \\
642 & $V_{6J}     $        &  $J^2(1-\cos 6\alpha)                                                $ &  & $  0.1254(10)\times 10^{-4}   $   \\
642 & $V_{6K}     $        &  $J_z^2(1-\cos 6\alpha)                                              $ &  & $ -0.1634(44)\times 10^{-3}   $    \\
642 & $V_{6zx}    $        &  $\frac{1}{2}\lbrace J_z,J_x \rbrace(1-\cos 6\alpha)                 $ &   & $ -0.752(33)\times 10^{-4}   $     \\
633 & $\rho_{3xy} $        &  $\frac{1}{2}\lbrace J_z,J_x,J_y,p_\alpha,\sin 3\alpha \rbrace       $ &   & $ 0.427(12)\times 10^{-5}   $      \\
624 & $V_{3JJ}    $        &  $J^4(1-\cos 3\alpha)                                                $ &   & $ -0.632(57)\times 10^{-10}   $    \\
624 & $V_{3JK}    $        &  $J^2J_z^2(1-\cos 3\alpha)                                           $ &  & $  -0.506(34)\times 10^{-8}   $    \\
624 & $F_{FJK}    $        &  $J^2J_z^2p_\alpha^2                                                 $ &  & $  -0.1344(47)\times 10^{-8}   $   \\
624 & $F_{KK}     $        &  $J_z^4p_\alpha^2                                                    $ &  & $  0.1582(46)\times 10^{-8}   $    \\
624 & $D_{3xyK}   $        &  $\frac{1}{2}\sin 3\alpha \lbrace J_z^2,J_x,J_y \rbrace               $ &                            $     $ & $ -0.967(40)\times 10^{-7}   $    \\
615 & $\rho_{JK}  $        &  $J^2J_z^3p_\alpha                                                   $ &   & $  0.382(32)\times 10^{-10}   $    \\
606 & $\Phi_{JK}  $        &  $J^4J_z^2                                                          $ &  & $  0.1475(33)\times 10^{-11}   $    \\
606 & $\Phi_{KJ}  $        &  $J^2J_z^4                                                          $ &   & $  -0.337(17)\times 10^{-11}   $    \\
606 & $\phi_{JK}  $        &  $J^2\lbrace J_z^2,(J_x^2-J_y^2) \rbrace                             $ &   & $ 0.820(27)\times 10^{-12}   $      \\
\hline          
\multicolumn{3}{c}{Number of parameters}  & 17  & 40 \\
\multicolumn{3}{c}{Number of lines v$_t$=0, v$_t$=1}     & 286, 0  & 1428, 321\\
\multicolumn{3}{c}{$F_{max }$ in GHz}     & 110 & 650 \\
\multicolumn{3}{c}{$J_{max}, K_{a,max}$ }     & 15, 10 & 95, 20 \\
\multicolumn{3}{c}{rms in kHz}   & 107 & 43.4 \\
\multicolumn{3}{c}{wrms unitless}   & 0.71 & 0.90 \\
\hline\hline
\end{tabular*}
\flushleft
\begin{enumerate}[nolistsep]
\item[$^{\mathrm{a}}$] $n = t + r$, where $n$ is the total order of the operator, $t$ is the order of the torsional part and $r$ is the order of the rotational part, respectively.
\item[$^{\mathrm{b}}$] Parameter nomenclature based on the subscript procedures of \citet{Xu2008305}. \item[$^{\mathrm{c}}$]$\lbrace A,B,C \rbrace = ABC+CBA$, $\lbrace A,B \rbrace = AB+BA$. The product of the operator in the third column of a given row and the parameter in the second column of that row gives the term actually used in the torsion-rotation Hamiltonian of the program, except for $F$, $\rho$ and $A_{RAM}$, which occur in the Hamiltonian in the form $F(p_\alpha-\rho P_a)^2+A_{RAM}P_a^2$. 
\item[$^{\mathrm{d}}$] The parameters obtain with a refit of the data from \citet{2019ECS.....3.1537M}. 
\item[$^{\mathrm{e}}$] fixed to the value from XIAM fit (see table 1) in \citet{2019ECS.....3.1537M}.
\item[$^{\mathrm{f}}$]All values are in $\mathrm{cm}^{-1}$ (except $\rho$ which is unitless). Statistical uncertainties are shown as one standard uncertainty in the units of the last two digits.
\end{enumerate}
\end{table*}
Spectral predictions for astronomical use were calculated using the set of RAM Hamiltonian parameters presented in last column of Table~\ref{tabpar}. The predictions include the rotational transitions of the ground and first excited torsional states of \ce{CH3CSNH2} calculated at 300~K for $J$ values up to 95 and in the frequency range up to 400~GHz. In addition, we provided the tabulated values for the partition function in Table~\ref{part_funct}, with $Q_\mathrm{tot}(T) = Q_\mathrm{vib}(T) \times Q_\mathrm{tr}(T)$. The torsion-rotational partition function $Q_\mathrm{tr}(T)$ was calculated from first principles, that is, via direct summation over the rotational-torsional states. 
In the presence of low-energy vibrationally excited states, the vibrational part of the partition function should be also taken into account to derive column densities. The vibrational partition function was calculated with respect to the zero-point energy level using the following expression: 

\begin{equation}
Q(T)_\mathrm{vib}=\prod\limits_{\substack{i=1}}^{3N-6} \frac{1}{1-e^{-E_i/kT}}.
\end{equation}

The normal modes were obtained from the harmonic force field ab initio calculation at the B3LYP/6-311G++(3dp, 2pd) level of theory and basis set using \texttt{Gaussian 09} \citep{g09}. The harmonic frequencies are given in Table~\ref{vib_freqs}. Only the seven lowest vibrational excited state levels were considered (mode 2 to 8 in the Table~\ref{vib_freqs}). The contribution of the torsional mode is included in $Q_\mathrm{tr}(T)$ and thus, it is not considered for $Q_\mathrm{vib}$. The remaining modes above $1000~\mathrm{cm}^{-1}$ were found to have no influence on the partition function calculations for the temperatures up to 300~K. The results of the fit, as well as the predictions, are available in the Supplementary Material associated with this article.

\begin{table}
\centering
\caption{Harmonic vibrational frequencies up to $1000~\mathrm{cm}^{-1}$ of \ce{CH3CSNH2} determined at the B3LYP/6-311++G(3df,2pd) level of theory and basis set}
\begin{tabular*}{0.6\columnwidth}{@{\extracolsep{\fill}}cc}
\hline \hline
Mode    &       Frequency       \\
                &       ($\mathrm{cm}^{-1}$)     \\
\hline 
1               &        36.0932    \\
2               &       347.4387    \\
3               &       378.6060    \\
4               &       426.8825    \\
5               &       518.6759    \\
6               &       613.2849    \\
7               &       732.7576    \\
8               &       987.8108    \\
9               &       1021.2864   \\
10              &       1036.7852   \\
11              &       1322.0596   \\
12              &       1377.2398   \\
13              &       1403.9333   \\
14              &       1482.4055   \\
15              &       1489.0705   \\
16              &       1637.4358   \\
17              &       3020.4774   \\
18              &       3071.0614   \\
19              &       3164.1245   \\
20              &       3564.4275   \\
21              &       3696.1525   \\
\hline \hline
\end{tabular*}
\label{vib_freqs}
\end{table}

\begin{table}
 \caption{Rotational, vibrational, and total partition functions of \ce{CH3CSNH2} at various temperatures}
\begin{center}
 \label{part_funct}
\begin{tabular*}{\columnwidth}{@{\extracolsep{\fill}}lrcr}
\hline\hline
 $T (K)$ & $Q(T)_\mathrm{tr}$ & $Q(T)_\mathrm{vib}$ & $Q(T)_\mathrm{tot}$ \\
\hline
300     &459047.9395     &2.0239  & 929054.6665 \\
220     &242388.0230      &1.3800  & 334515.0879 \\
170     &141743.4841     &1.1541  & 163585.3842 \\
150     &109181.0704      &1.0949  & 119537.4570 \\
70      &22675.8793     &1.0014  & 22707.4107  \\
40     &7672.9688      &1.0000 &  7673.0084 \\
20     &2332.7909      &1.0000  &  2332.7909  \\
10     &784.3603      &1.0000  & 784.3603  \\
\hline
\end{tabular*}
\end{center}
\end{table}

\section{Expanded astronomical searches \label{Sec:Observation}}

In the initial analysis characterizing the spectroscopy of \ce{CH3CSNH2}, two astronomical sources were selected from the ASAI survey to search for transitions of \ce{CH3CSNH2} based on the initial fit from that work \citep{2019ECS.....3.1537M}. The ASAI survey targets sources associated with Solar-type star forming regions over a wide range of evolutionary states, temperatures, densities and kinematics providing a diverse sample of sources to investigate \citep{2018MNRAS.477.4792L}. The protostellar core L1544 and the shocked region L1157-B1 associated with the chemically rich outflow L1157 were the two sources that were selected. Both regions are spatially extended (with respect to the IRAM 30m beam). No transitions of \ce{CH3CSNH2} were detected above the $3\sigma$ noise level of those observations. That study assumed a dipole moment of $\sim4\,\mathrm{D}$ in order to calculate the upper limit to the column density of \ce{CH3CSNH2} toward these sources. For both sources, the upper limit to the beam averaged column density was $<10^{12}~\mathrm{cm}^{-2}$. With the improved spectroscopic observations of this study and the inclusion of the vibrational part of the partition function, it is now possible to perform a more extensive search for \ce{CH3CSNH2} and to provide a more robust constraint on column densities.

\subsection{Search for \ce{CH3CSNH2} toward Sgr B2(N2) with ALMA\label{Sec:sgrb2n2}}

We searched for \ce{CH3CSNH2} toward the hot molecular core Sgr B2(N2). This source belongs to the protocluster Sgr~B2(N) that contains several hot cores and (ultracompact) HII regions \citep[for example,,][]{2017A&A...604A..60B,1995ApJ...449..663G}, as well as a large number of compact continuum sources detected with the Atacama Large Millimeter/submillimeter Array (ALMA) that are characterized by densities above $10^8~\mathrm{cm}^{-3}$ and were interpreted as being high mass young stellar objects \citep[][]{2017A&A...604A...6S,2018ApJ...853..171G}. Sgr~B2(N) has often been targeted over the past five decades to search for complex organic molecules thanks to the high column densities of its hot cores that facilitate the detection of these molecules.

We used the data from the Exploring molecular complexity with ALMA (EMoCA) imaging spectral line survey performed with ALMA in Cycles 0 and 1 to search for \ce{CH3CSNH2} toward Sgr~B2(N2). Details about the observations and data reduction of this survey can be found in \citet[][]{2016A&A...587A..91B}. In short, the survey covers the frequency range from 84.1 to 114.4~GHz with a spectral resolution of 488~kHz that corresponds to 1.7 to 1.3 $\mathrm{km}$ $\mathrm{s^{-1}}$ in velocity space. The median angular resolution of the survey is $1.6\arcsec$ which corresponds to $\sim$13000~au at a distance of 8.15~kpc \citep[][]{2019ApJ...885..131R}. The phase center of the interferometric observations was located at ($\alpha, \delta$)$_{\rm J2000}$=($17^{\rm h}47^{\rm m}19{\fs}87, -28^\circ22'16{\farcs}0$). The spectrum that is analyzed here corresponds to the peak position of Sgr~B2(N2) located at ($\alpha, \delta$)$_{\rm J2000}$=($17^{\rm h}47^{\rm m}19{\fs}86$, $-28^\circ22\arcmin13{\farcs}4$). 

We used the software \texttt{WEEDS} \citep[][]{2011A&A...526A..47M} to model the observed spectrum under the assumption of local thermodynamic equilibrium (LTE) and produce a synthetic spectrum. This assumption is justified by the high densities that characterize the region where the hot core emission is detected ($>1 \times 10^7\ \mathrm{cm^{-3}}$, see \citealt{2019A&A...628A..27B}). As described in \citet{2016A&A...587A..91B}, we modeled the emission of each molecule separately and then added their contributions together. We modeled the spectrum of each species with five parameters: size of the emitting region, column density, temperature, line width, and velocity offset with respect to the assumed systemic velocity of the source ($74\ \mathrm{km\ s}^{-1}$).

\begin{figure}
\centerline{\resizebox{0.98\hsize}{!}{\includegraphics[angle=0]{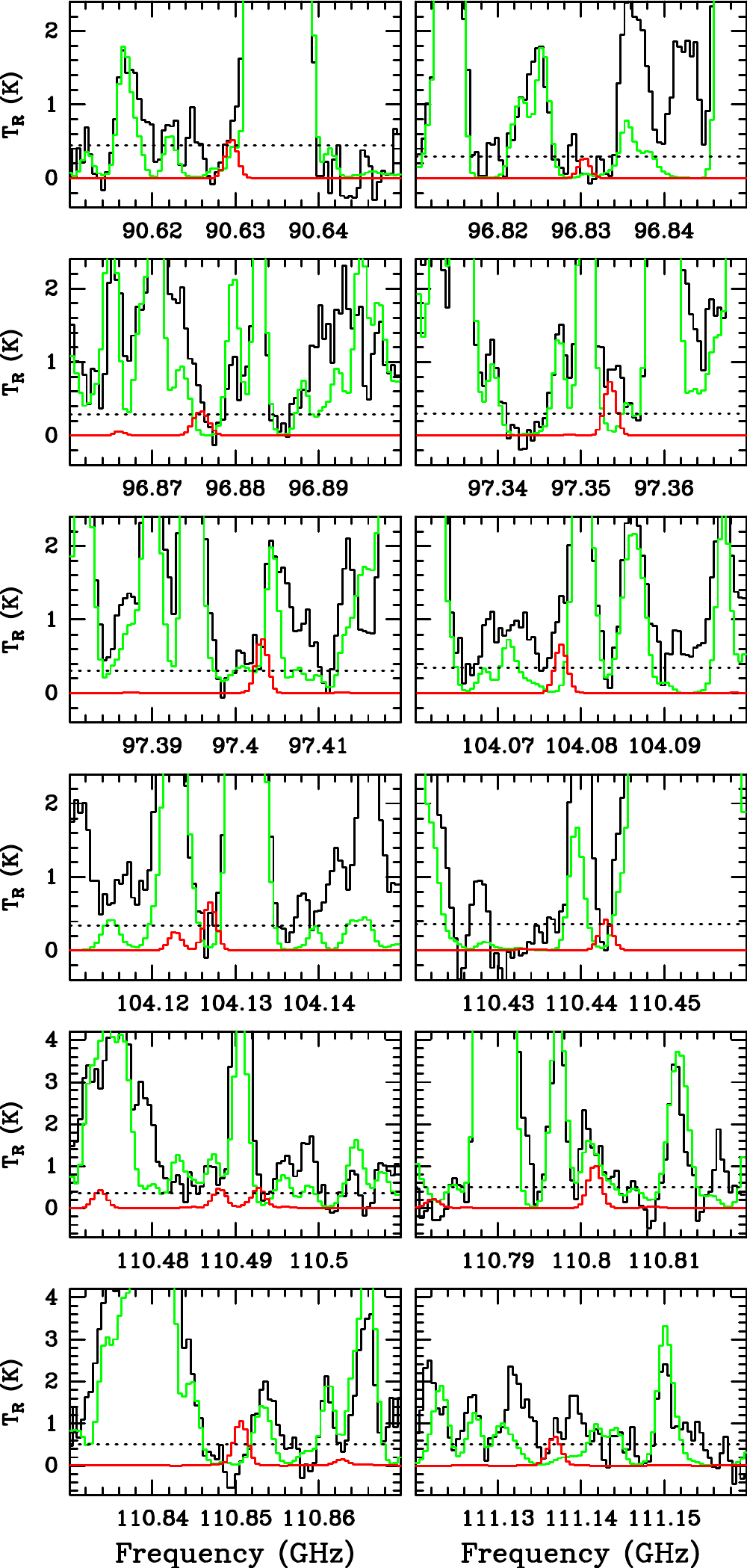}}}
\caption{Synthetic LTE spectrum of \ce{CH3CSNH2} (in red) used to derive the upper limit to its column density, overlaid on the ALMA spectrum of Sgr~B2(N2) (in black) and the synthetic spectrum that contains the contributions of all the species (but not \ce{CH3CSNH2}) that we have identified so far in this source (in green). The dotted line in each panel indicates the $3\sigma$ noise level. The transitions of \ce{CH3CSNH2} that have a modeled peak temperature lower than 3$\sigma$ and the ones that are heavily contaminated by other species are not shown.}
\label{f:spec_ch3csnh2_ve0}
\end{figure}

We did not detect emission from \ce{CH3CSNH2} toward Sgr~B2(N2) with the EMoCA survey. In order to derive an upper limit to its column density, we assumed that its emission should trace a similar region as \ce{CH3CONH2} and we fixed the size of the emitting region, temperature, line width, and velocity offset to the values derived for \ce{CH3CONH2} in Sgr~B2(N2) by \citet{2017A&A...601A..49B}. The only free parameter was then the column density of \ce{CH3CSNH2}. Figure~\ref{f:spec_ch3csnh2_ve0} shows the synthetic spectrum of \ce{CH3CSNH2} that we used to derive the upper limit to its column density. This upper limit is reported in Table~\ref{t:coldens}, along with the parameters derived earlier for \ce{CH3CONH2}, methanol (\ce{CH3OH}), and methyl mercaptan (\ce{CH3SH}) by \citet{2017A&A...601A..49B} and \citet{2016A&A...587A..92M} from the EMoCA survey.

\input{tab_thioacetamide_weedsmodel.tex}

Table~\ref{t:coldens} indicates that \ce{CH3CSNH2} is at least $\sim9$ times less abundant than \ce{CH3CONH2} in Sgr~B2(N2). For comparison, \ce{CH3SH} is $\sim120$ times less abundant than \ce{CH3OH} in this source. If these two pairs of molecules have similar abundance ratios, then the emission lines of \ce{CH3CSNH2} may be one order of magnitude weaker than the upper limits we obtained with the EMoCA survey, and thus most likely well below the spectral confusion limit toward this source in this frequency range.

\subsection{Search for \ce{CH3CSNH2} toward Sgr B2(N) with the GBT \label{Sec:sgrb2-primos}}

The PRebiotic Interstellar Molecule Survey (PRIMOS) \footnote{Access to the entire PRIMOS data set, specifics on the observing strategy, and overall frequency coverage information is available at \url{http://archive.nrao.edu} by searching for GBT Program ID: AGBT07A\_051} was a key science program that started in 2008 January and concluded in 2011 July taken with the Robert C. Byrd Green Bank Telescope (GBT) currently managed by the Green Bank Observatory (GBO). The PRIMOS project covers nearly all observable frequencies available to ground based instrumentation from $\sim300\ \mathrm{MHz}$ to 48~GHz at high sensitivity ($\sim~3-9\ \mathrm{mK}$ rms) and spectral resolution (24.4~kHz). The pointing position for these observations were centered on the Sgr B2(N) Large Molecule Heimat (LMH) at ($\alpha, \delta$)$_{\rm J2000}$=($17^{\rm h}47^{\rm m}19{\fs}8, -28^\circ22'17{\farcs}0$), nearly identical to the pointing position of the EMoCA survey (see Section~\ref{Sec:sgrb2n2}).

The systemic velocities found for \ce{CH3CONH2} in the original detection paper \citep{2006ApJ...643L..25H} are +64, +73 and +82 $\mathrm{km\ s}^{-1}$.  In general, there is much less contribution from the $+73 \mathrm{km\ s}^{-1}$ component compared to $+64\ \mathrm{km\ s}^{-1}$ and $+82\ \mathrm{km\ s}^{-1}$ in the GBT observations.  However, the ALMA observations are also sensitive to the compact emission regions of molecular gas (the resolution and associated source sizes from the EMoCA survey are $\sim1.6\arcsec$ whereas the GBT observations are sensitive to a range of resolutions and source sizes, dependent on observing frequency, from $\sim15\arcsec$ to $\sim80\arcsec$). As such, from these two complementary data sets, it is possible to investigate the hot compact regions as well as the colder more extended molecular gas in searching for transitions of \ce{CH3CSNH2} toward the Sgr B2(N) region. Intensities are presented on the $T_\mathrm{A}^{*}$ scale \citep{1976ApJS...30..247U}. 

The detection of large molecules with the PRIMOS data set for the last 15 years has come as a surprise as previously, the prevailing theories for the formation of large molecules was believed to be driven primarily by grain surface chemistry during a warm-up phase during the process of star formation \citep{2008ApJ...682..283G}.  As such, the regions surrounding hot molecular cores would be ideal for these discoveries (See Section~\ref{Sec:sgrb2n2}).  Yet the PRIMOS observations have shown that there is a large diversity of complex molecules in the clouds surrounding the Sgr B2(N) region and not specifically centered on the hot core regions.  In addition, PRIMOS observations have also shown a rich and complex molecular diversity among the spiral arms clouds in the intervening gas \citep{2018A&A...610A..10C} toward the Galactic Center.  These observations together indicate that gas phase chemistry may drive the formation of molecules that cannot happen on grain surfaces \citep{2012ApJ...755..153N}.  Even more interesting was the conclusion that most transitions of these larger molecules (for example, \ce{CH3OCHO} and \ce{CH2NH}) with energy levels below 30 GHz are astronomical masers \citep{2014ApJ...783...72F, 2018JPCL....9.3199F} leading to the detection of new molecules which would have otherwise eluded detection \citep{2012ApJ...758L..33M}.  

Since the initial observations from the first detection of \ce{CH3CONH2} \citep{2006ApJ...643L..25H}, additional transitions have been observed with the GBT as part of the PRIMOS survey.  Also, the initial detection did not include any of the hyperfine structure (HFS) in the modeling of the spectral line profiles from PRIMOS nor in the determination of the total column density.  In this work, spectral line profiles of the low frequency transitions were modeled for all the A-state transitions below 48 GHz.  In addition, transitions below 18 GHz were modeled using the measured HFS (F. J.\ Lovas, private communication).  Above 18 GHz, the HFS is no longer resolved at the resolution of our astronomical observations. We used the software \texttt{MOLSIM} \citep{Lee_molsim_2021} to perform the modeling and to estimate the upper limit for the spectra analysis of the PRIMOS and ASAI surveys. Synthetic spectra are produced with a single excitation temperature assumption, that all transitions are thermalized, and corrected for optical depth as discussed in \citet{1991ApJS...76..617T} and \citet{2015PASP..127..266M}.

\begin{figure}[t!]
\centerline{\resizebox{0.98\hsize}{!}{\includegraphics[angle=0]{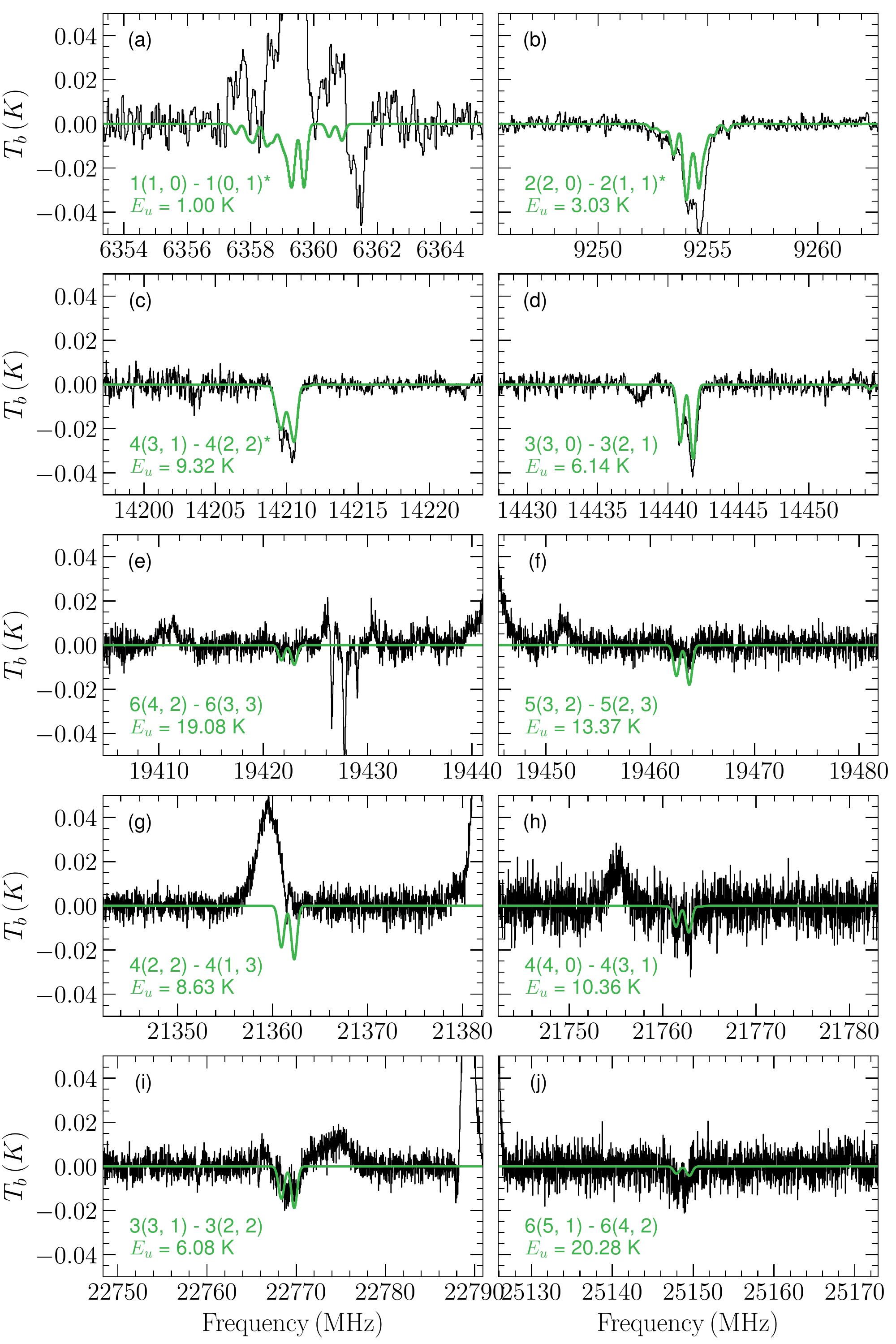}}}
\caption{Synthetic LTE spectrum of \ce{CH3CONH2} (in green) A-symmetry state transitions generated from the physical conditions and observing parameters described in Section~\ref{Sec:sgrb2-primos} ($\mathrm{T_{ex}} = 5.8$ K,  source size of $20\arcsec$, three velocity components at +62, +71 and $+81 \ \mathrm{km\ s}^{-1}$ yielding an $\mathrm{N_T}\sim7.7\times10^{13}\ \mathrm{cm}^{-2}$ from all three components) overlaid on the GBT spectrum of Sgr B2(LMH) (in black).  Transition quantum numbers and upper state energy levels (Table~\ref{tab:acet_spec_dat}) are in the bottom left corner of each spectrum.  Quantum numbers with an "$\star$" have HFS included in the synthetic spectrum.}
\label{f:Acetamide_A_States_PRIMOS}
\end{figure}

\begin{figure}[h!]
\centerline{\resizebox{0.78\hsize}{!}{\includegraphics[angle=0]{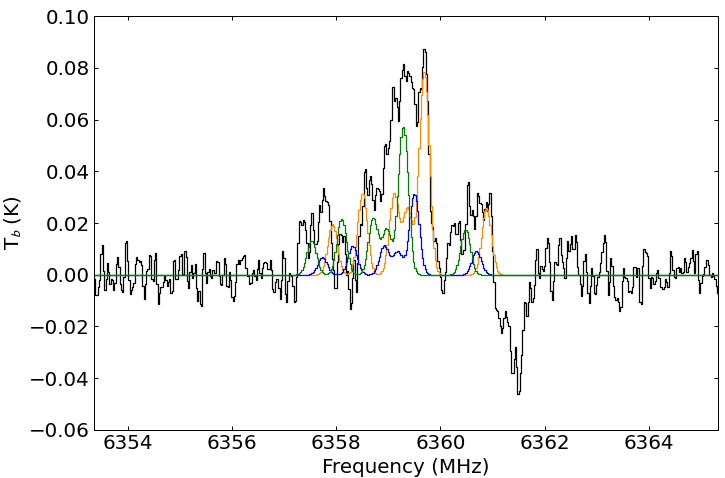}}}
\caption{Illustrative spectrum of \ce{CH3CONH2} showing the relative intensities of all the HFS (Table~\ref{tab:acet_spec_dat}) and multiple velocity components at +62 (in yellow), +71 (in blue) and $+81\ \mathrm{km\ s}^{-1}$ (in green) overlaid on the GBT spectrum of Sgr B2(LMH) (in black).  The intensity of these emission features cannot be matched by the synthetic LTE model in this work.}
\label{f:HFS_Vel_Maser_PRIMOS}
\end{figure}

Acetamide (\ce{CH3CONH2}) is an oblate, asymmetric top molecule ($K_c$ is a good quantum number) with an internal methyl rotor \citep{2002JMoSp.215..144Y}. The dipole moments of \ce{CH3CONH2} are $\mu_a = 1.14 D$ and $\mu_b =3.5 D$ \citep{2006ApJ...643L..25H}.  As such,  b-type transitions would be inherently stronger than a-type transitions.  Several attributes of the detected transitions in PRIMOS were immediately found.  First, as expected, only b-type transitions were detected.  Second, only $\Delta J = 0$, Q-branch transitions were detected with $\Delta K_a = 1$ and $\Delta K_c = -1$ for the A-symmetry states (hereafter, the convention $^b$Q$_{1,-1}$ will be used to describe the types of transitions).  Figure~\ref{f:Acetamide_A_States_PRIMOS} shows all the Q-branch, A-symmetry state, b-type transitions observed in the PRIMOS survey.  The lowest frequency transition detected near 6358 MHz (Figure~\ref{f:Acetamide_A_States_PRIMOS}a) shows resolved HFS, multiple velocity components at +62, +71 and $+81\ \mathrm{km\ s}^{-1}$ and maser activity.  The model fit (red trace) cannot reproduce the measured emission features detected from this transition.  In contrast, nearly all other transitions detected are well characterized by the LTE model - even the weaker features.  The transition near 9253 MHz (Figure~\ref{f:Acetamide_A_States_PRIMOS}b) has a slightly stronger $+62 \mathrm{km\ s}^{-1}$ absorption component than predicted by the model.  This can be explained by possible contamination from other spectral features near those frequencies in either absorption or emission.  For example, the H(113)$\epsilon$ recombination spectral line profile is clearly impacted by absorption due to the acetamide transition near 21362 MHz (Figure~\ref{f:Acetamide_A_States_PRIMOS}g).  What also differs from the initial detection paper by \citet{2006ApJ...643L..25H} is that the best fit velocity components of \ce{CH3CONH2} occur at +62, +71 and $+81 \ \mathrm{km\ s}^{-1}$ and not at the notional +64, +73 and $+82 \ \mathrm{km\ s}^{-1}$ identified in other species - suggesting a spatial differentiation of acetamide from other species such as formamide, glycolaldehyde, propynal and cyclopropenone \citep{2004ApJ...610L..21H, 2004ApJ...613L..45H, 2006ApJ...642..933H, 2013A&A...559A..47B}. Table~\ref{tab:acet_spec_dat} presents the spectroscopic parameters for the \ce{CH3CONH2} transitions detected in this work including the HFS for frequencies below 18 GHz (where resolved) for the A-symmetry states and Figure~\ref{f:HFS_Vel_Maser_PRIMOS} shows the 6358 MHz transition illustrating all the HFS and velocity components.

\begin{table*}
    \centering
    \caption{Spectroscopic properties of \ce{CH3CONH2} lines detected in PRIMOS toward Sgr B2(LMH).}
    \begin{tabular}{ccccrrc}
    \hline\hline
    \multicolumn{2}{c}{Transitions} &{Symm.} &{Frequency} &{$E_{up}$} &{$S_{ij}\mu^2$} & Transition \\
    {$J'(K_a',K_c')\ - \ J''(K_a'',K_c'')$} & {$F'\ -\ F''$} & & {(MHz)} & {(K)} & {(D$^2$)} & Type\\
    \hline
1(1,0) - 1(0,1) & F = 1-1 & A & 6357.895(8) & 0.9988  & 1.595 & $^b$Q$_{1,-1}$ \\
                & F = 1-2 & A & 6358.472(7) & 0.9988  & 2.658 & \\
                & F = 2-1 & A & 6359.078(6) & 0.9989  & 2.657 & \\
                & F = 1-0 & A & 6359.338(8) & 0.9988  & 2.126 & \\
                & F = 2-2 & A & 6359.656(5) & 0.9989  & 7.973 & \\
                & F = 0-1 & A & 6360.853(8) & 0.9990  & 2.125 & \\
                
2(1,1) - 1(1,0) & & E & 9050.672(1) & 8.8659 & 4.154 & $^a$P$_{0,1}$\\

2(2,0) - 2(1,1) & F = 2-1 & A & 9252.990(7) & 3.0293  & 1.247 & $^b$Q$_{1,-1}$ \\
                & F = 2-3 & A & 9253.351(6) & 3.0293  & 1.294 & \\
                & F = 2-2 & A & 9254.000(6) & 3.0293  & 5.771 & \\
                & F = 3-3 & A & 9254.535(4) & 3.0293  & 10.339 & \\
                & F = 1-1 & A & 9254.833(6) & 3.0294  & 3.738 & \\
                & F = 3-2 & A & 9255.185(6) & 3.0293  & 1.293 & \\
                & F = 1-2 & A & 9255.843(7) & 3.0294  & 1.246 & \\
                
1(1,1) - 1(0,1) & & E &  13388.703(1) & 7.4374 & 7.644 & $^a$Q$_{1,0}$ \\

4(3,1) - 4(2,2) & F = 4-3 & A & 14209.211(10) & 9.3165 & 0.749 & $^b$Q$_{1,-1}$ \\
                & F = 4-5 & A & 14209.393(9) & 9.3165 & 0.753 & \\
                & F = 4-4 & A & 14210.101(7) & 9.3165 & 13.901 & \\
                & F = 5-5 & A & 14210.440(7) & 9.3165 & 18.072 & \\
                & F = 3-3 & A & 14210.527(7) & 9.3165 & 11.233 & \\
                & F = 5-4 & A & 14211.147(9) & 9.3165 & 0.753 & \\
                & F = 3-4 & A & 14211.416(10) & 9.3165 & 0.749 & \\

3(3,0) - 3(2,1) & & A & 14441.706(1) & 6.1401 & 24.436 & $^b$Q$_{1,-1}$ \\
                
                
2(1,2) - 1(1,1) & & E &  14651.890(1) & 8.1405 & 11.226 & $^a$P$_{0,1}$ \\
                
3(1,2) - 3(1,3) & & E &  14960.874(1) & 10.4612 & 4.369 & $^a$Q$_{0,-1}$ \\

2(1,1) - 2(1,2) & & E &  15115.748(1) & 8.8659 & 13.673 & $^a$Q$_{0,-1}$ \\

4(3,2) - 4(2,2) & & E &  18507.373(2) & 15.0377 & 23.916 & $^b$Q$_{1,0}$ \\

6(4,2) - 6(3,3) & & A &  19422.841(1) & 19.0778 & 65.746  & $^b$Q$_{1,-1}$ \\

5(3,2) - 5(2,3) & & A & 19463.627(1) & 13.3666 & 49.517  & $^b$Q$_{1,-1}$ \\

3(2,1) - 3(2,2) & & E & 20686.104(1) & 12.0580 & 23.052 & $^a$Q$_{0,-1}$ \\

2(2,1) - 2(1,1) & & E & 20891.602(1) & 9.8686 & 16.739 & $^b$Q$_{1,0}$ \\

4(2,2) - 4(1,3) & & A & 21362.137(1) & 8.6346 & 30.887  & $^b$Q$_{1,-1}$ \\

4(4,0) - 4(3,1) & & A & 21762.725(2)  & 10.3609 & 22.240  & $^b$Q$_{1,-1}$ \\

2(0,2) - 1(0,1) & & E & 22095.527(1) & 7.8551 & 18.829 & $^a$P$_{0,1}$ \\

3(3,1) - 3(2,2) & & A & 22769.640(1) & 6.0783 & 16.040  & $^b$Q$_{1,-1}$ \\

6(5,1) - 6(4,2) & & A & 25149.383(2) & 20.2848 & 45.460  & $^b$Q$_{1,-1}$ \\

4(3,1) - 4(3,2) & & E & 26196.049(1) & 16.2949 & 30.516 & $^a$Q$_{0,-1}$ \\

\hline
    \end{tabular}
   \tablefoot{\ce{CH3CONH2}  rest  frequencies  from  \citet{2004JMoSp.227..140I} and F. J.\ Lovas, private communication.  Uncertainties  in  parentheses refer to the least significant digit and are 2$\sigma$ values (type A coverage) \citep{JPCS_772_1_012024bib3}.}
\label{tab:acet_spec_dat}
\end{table*}

\begin{figure}
\centerline{\resizebox{0.98\hsize}{!}{\includegraphics[angle=0]{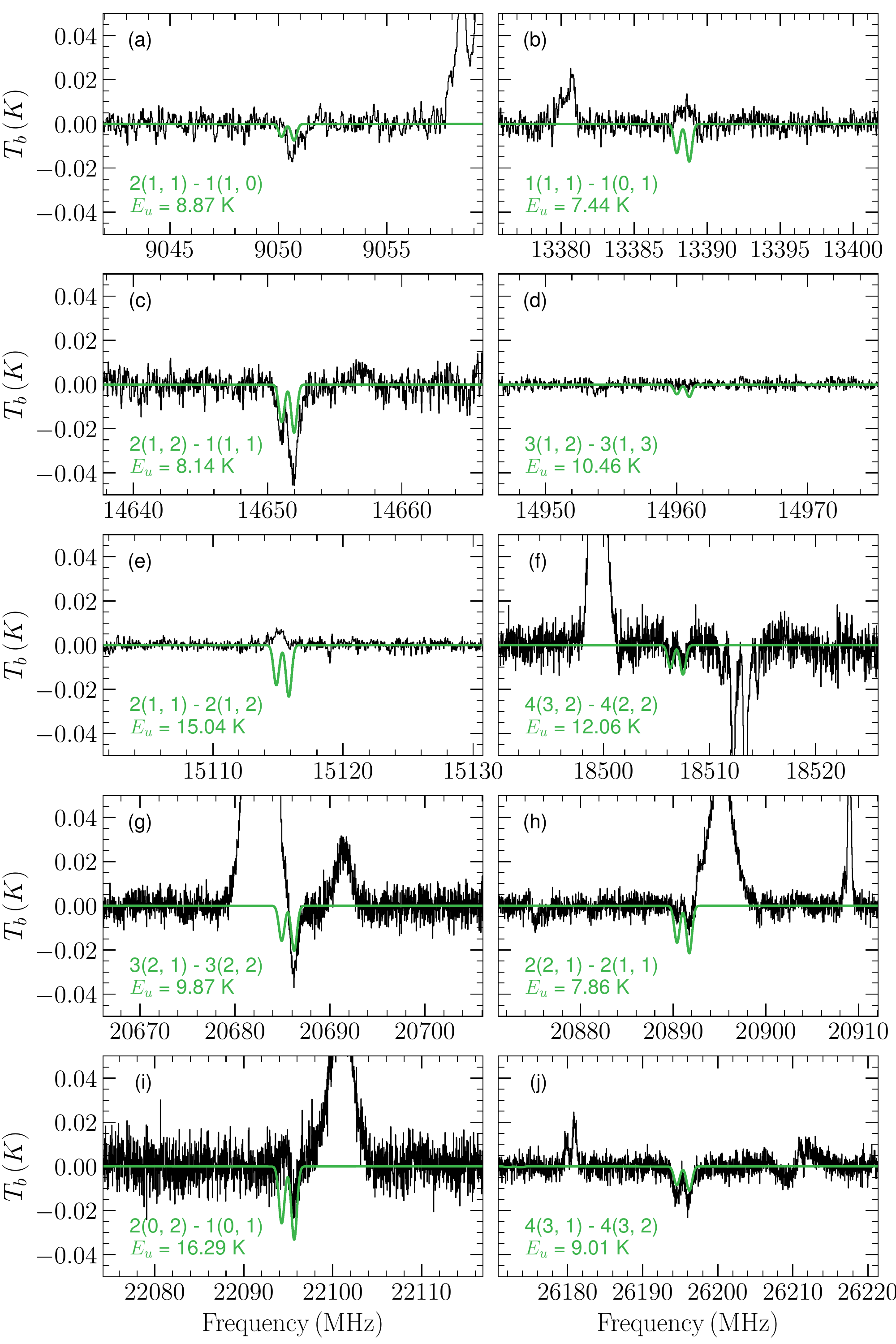}}}
\caption{Synthetic LTE spectrum of \ce{CH3CONH2} (in green) E-symmetry state transitions generated from the physical conditions and observing parameters described in Section~\ref{Sec:sgrb2-primos} ($\mathrm{T_{ex}}$ = 5.8 K,  source size of $20\arcsec$, three velocity components at +62, +71 and $+81 \ \mathrm{km\ s}^{-1}$ yielding an $\mathrm{N_T}\sim1.5\times10^{14}\ \mathrm{cm}^{-2}$ from all 3 components) overlaid on the GBT spectrum of Sgr B2(LMH) shown in black.  Transition quantum numbers and upper state energy levels (Table~\ref{tab:acet_spec_dat}) are in the bottom left corner of each spectrum.}
\label{f:Acetamide_E_States_PRIMOS}
\end{figure}

The situation is not as clear for the E-symmetry state transitions.  Figure~\ref{f:Acetamide_E_States_PRIMOS} shows the detection of ten E-state transitions.  Unlike the A-state transitions, given the difficulty in fitting the E-state, there is no corresponding HFS data available.  In addition, the astronomically detected E-state transitions in the low frequency data are not limited to just $^b$Q$_{1,-1}$ type transitions.  For example, the feature detected at 26196 MHz (Figure~\ref{f:Acetamide_E_States_PRIMOS}j) is a  $^a$Q$_{0,-1}$ transition; the features detected at 9050 (Figure~\ref{f:Acetamide_E_States_PRIMOS}a) and 14652 MHz (Figure~\ref{f:Acetamide_E_States_PRIMOS}c) are $^a$P$_{0,1}$ transitions; finally the feature detected at 13388 MHz (Figure~\ref{f:Acetamide_E_States_PRIMOS}b) is a  $^a$Q$_{0,1}$ transition that appears to be a weak maser transition.  Nonetheless, the best fit parameters found for the E-state transition are constrained to be similar to that for the A-states; that is the model fits the data very well for an excitation temperature similar to what was determined for the A-state but with a factor of $\sim$2 times higher column density for those features found in absorption and not contaminated by other spectral features.  Similar to the detected A-state transitions, several E-state transitions are contaminated by features from H-recombination lines (H(104)$\delta$ at 22095 MHz, H(106)$\delta$ at 20891 MHz, and H(85)$\beta$ at 20686 MHz) or possibly other unidentified transitions. We note that the 15115 MHz \ce{CH3CONH2} transition (Figure~\ref{f:Acetamide_E_States_PRIMOS}e) only shows emission near the 71 km/s velocity component and no other features at 62 or 81 km/s suggesting that this feature is also a maser transition or possibly contaminated by another species.  The largest discrepancy between the data and the fit is around 14960 MHz.  With no other transitions in the passband to interfere with the detection, this feature is markedly absent in the astronomical data compared to the other detected transitions.  This transition is a $^a$Q$_{0,-1}$ similar to the transitions shown in Figure~\ref{f:Acetamide_E_States_PRIMOS}e, \ref{f:Acetamide_E_States_PRIMOS}g, and \ref{f:Acetamide_E_States_PRIMOS}j.  Looking at the energy levels of the transitions in these figures, it appears that for transitions with upper state energies less than 10 K, features are detected in absorption (\ref{f:Acetamide_E_States_PRIMOS}g, and \ref{f:Acetamide_E_States_PRIMOS}j).  At 15 K, the feature of this transition type is in emission (\ref{f:Acetamide_E_States_PRIMOS}e).  And, at $\sim$11 K, features from this transition type are absent (\ref{f:Acetamide_E_States_PRIMOS}d).  It is possible that features from these types of transitions pass from emission (when $E_u \sim 15 \mathrm{K}$) to absorption (when $E_u$ $\sim$ 9.87 and 9.0 $\mathrm{K}$) passing through an upper state energy around 11 K.  This could account for the transition being absent compared to the model data.  The transition is also unique in that it is a  $J - K_a = 2$ transition for both the upper and lower states.  None of the other astronomically detected transitions are of this type.  Table~\ref{tab:acet_spec_dat} presents the spectroscopic data for the E-state \ce{CH3CONH2} transitions detected in the PRIMOS observations.

In total, the best fit model to the PRIMOS data was found for an excitation temperature of 5.8 K and a source size of $20\arcsec$ for both the A- and E-state transitions, and the total measured \ce{CH3CONH2} column density determined from the PRIMOS data from the current analysis, is $\sim7.7\times 10^{13}\ \mathrm{cm}^{-2}$ for the A-state transitions and $\sim1.5\times 10^{14}\ \mathrm{cm}^{-2}$ for the E-state transitions.  These values are on the same order of magnitude with both the $\sim8\times 10^{13}\ \mathrm{cm}^{-2}$ column density determined by \citet{2006ApJ...643L..25H} at an excitation temperature of 5 K and the $\sim5\times 10^{13}\ \mathrm{cm}^{-2}$ determined by \citet{2011ApJ...743...60H} at an excitation temperature of 17 K for transitions detected with upper state energy levels lower than 40 K.  And, as expected, the observed temperature and measured column densities from this current work are much lower than what was reported by \citet{2017A&A...601A..49B} toward SgrB2(N2) from the EMoCA survey where \ce{CH3CONH2} was detected at a column density of  $\sim2\times 10^{17}\ \mathrm{cm}^{-2}$ at an excitation temperature of 180 K (see Table~\ref{t:coldens})  - clearly emission coming from a hot core region which is beam diluted in our GBT observations.  What this analysis also indicates, given the structural similarity of \ce{CH3CONH2} and \ce{CH3CSNH2}, are the types of transitions that are most likely to be detected at lower frequencies with single dish instruments such as the GBT.

\begin{figure}[t!]
\centerline{\resizebox{0.98\hsize}{!}{\includegraphics[angle=0]{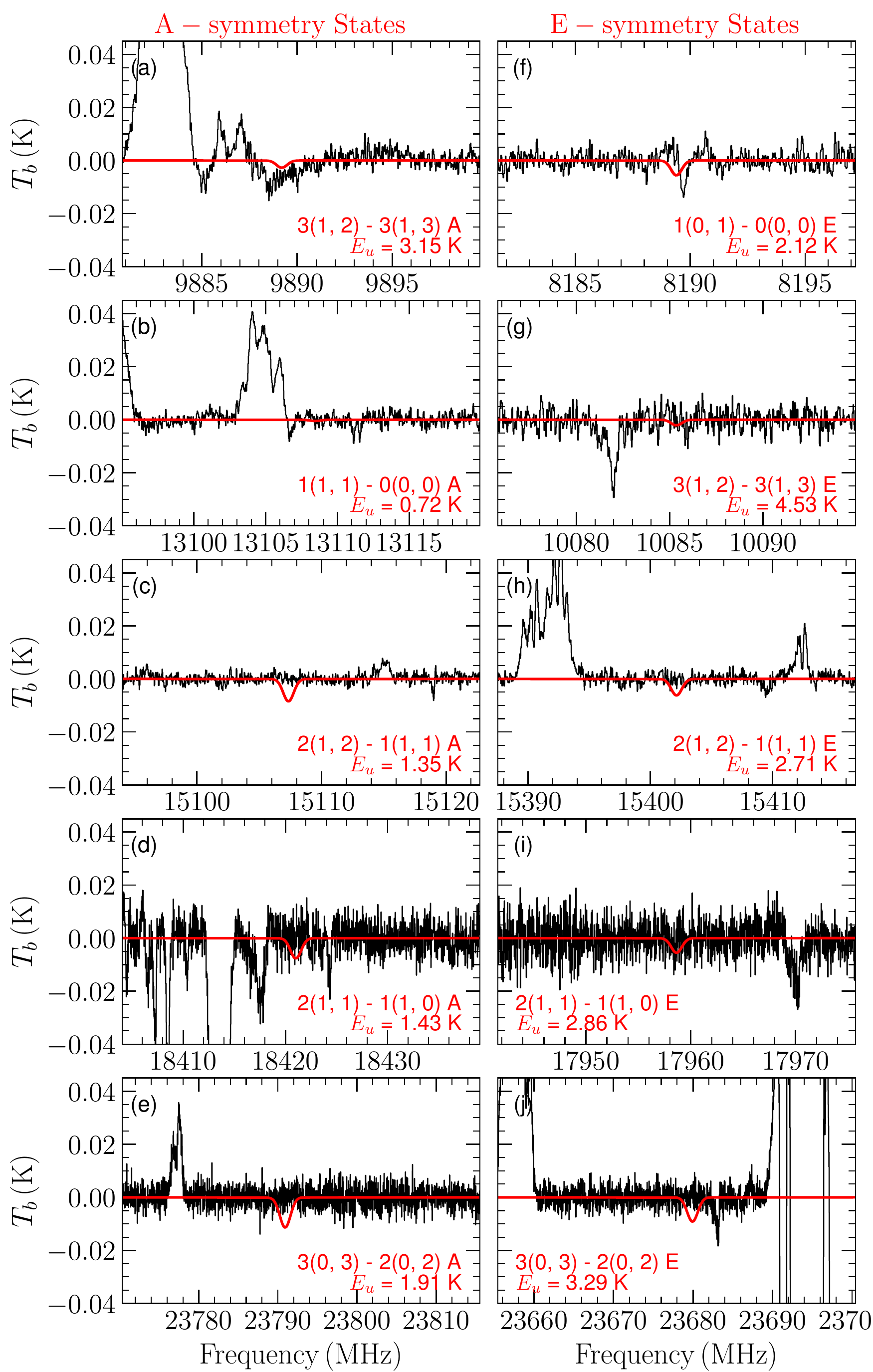}}}
\caption{Synthetic LTE spectrum of \ce{CH3CSNH2} (in red).  The A-symmetry state transitions are presented in the first column in panels (a)-(e) and the E-symmetry state transitions are presented in the second column in panels (f)-(j).  The spectrum was generated from the physical conditions and observing parameters described in Section~\ref{Sec:sgrb2-primos} (that is, T$_{ex}$ = 5.8,  source size of $20\arcsec$, $N_T\sim2.2\times10^{13} \ \mathrm{cm}^{-2}$) overlaid on the GBT spectrum of Sgr B2(LMH) shown in black.  Transition quantum numbers and upper state energy levels (Table~\ref{tab:thio_spec_dat}) are at the bottom of each spectrum.  No significant emission from \ce{CH3CSNH2} transitions are detected beyond the $3\sigma$ upper limit.}
\label{f:Thioacetamide_States_PRIMOS}
\end{figure}

Figure~\ref{f:Thioacetamide_States_PRIMOS} shows the results of our attempt to detect the strongest transitions, including possible maser transitions, from \ce{CH3CSNH2} between 8 and 24~GHz from the PRIMOS survey.  Since the fit to \ce{CH3CSNH2} was done utilizing the rho-axis method (see Section~\ref{Sec:analysis of the spectra}), A- and E-symmetry state transitions from \ce{CH3CSNH2} were both initially searched for, though due to the limited Doppler resolution used in the spectrometer, it was not possible to search for any HFS.  Also, given $\mu_a = 4.1 D$ for \ce{CH3CSNH2}, the strongest transitions should be a-type transitions ($\mu_b = -1.1 D$).  Finally, the lowest energy state transitions could also be possible maser transitions.  And, for completeness, we searched for all A- and E-symmetry state, P- and Q-branch, a- and b-type transitions over the entire PRIMOS observing range. 

As was done when first characterizing the astronomical spectra of \ce{CH3CONH2}, we used an LTE-model\footnote{As with \ce{CH3CONH2}, maser transitions are identified because they greatly deviate from the predicted LTE-model.  A full characterization of the spectrum including this maser activity is beyond the scope of this work.} assuming a non-thermal background source of emission \citep{2007ApJ...660L.125H, 2016Sci...352.1449M} and simulated the spectrum of \ce{CH3CSNH2} assuming a temperature of 5.8 K, a source size of $\sim20\arcsec$ and a line width of $20\ \mathrm{km\ s}^{-1}$.

However, as Figure~\ref{f:Thioacetamide_States_PRIMOS} illustrates, similar to what was found in the EMoCA survey, no \ce{CH3CSNH2} transitions (in red) were detected - although there is a nearly frequency coincident absorption feature close to the 1(0,1) - 0(0,0) E transition at 8189.41 MHz.  The lack of exact frequency coincidence and of any additional transitions detected however, makes this feature very unlikely to be due to \ce{CH3CSNH2}.   To set an upper limit of \ce{CH3CSNH2}, a rotational temperature of 5.8 K was used and based on the lowest noise level from the \ce{CH3CSNH2} passbands, an upper limit of $2.2\times10^{13} \ \mathrm{cm}^{-2}$ was found -  again indicating that \ce{CH3CSNH2} is at least an order of magnitude less abundant than \ce{CH3CONH2} in the low temperature, extended regions around Sgr~B2(N).

\begin{table*}
    \centering
    \caption{Spectroscopic properties of the \ce{CH3CSNH2} lines shown toward the Sgr B2 and NGC 6334I regions.}
    \begin{tabular}{lcrrrc}
    \hline\hline
    Transition & Symm. & Frequency & $E_{up}$ & $S_{ij}\mu^2$  & Transition \\
    {$J'(K_a',K_c')\ - \ J''(K_a'',K_c'')$} & & {(MHz)} & {(K)} & {(D$^2$)} & Type\\
    \hline
1(0,1) - 0(0, 0)& E & 8189.4099(78)     &1.80   &15.7   & $^a$P$_{0,1}$\\
3(1,2) - 3(1, 3)& A & 9889.2108(51)     &2.91   &9.3    & $^a$Q$_{0,-1}$\\
                & E & 10085.3720(121)   &4.30   &9.3    & $^a$Q$_{0,-1}$\\
1(1,1) - 0(0, 0)& A & 13108.4651(33)    &0.63   &1.2    & $^b$P$_{1,1}$\\
2(1,2) - 1(1, 1)& A & 15107.3571(9)     &1.35   &23.5   & $^a$P$_{0,1}$\\
                & E & 15402.1503(85)    &2.73   &22.1   & $^a$P$_{0,1}$\\
2(1,1) - 1(1, 0)& E & 17958.6594(97)    &3.01   &21.4   & $^a$P$_{0,1}$\\
                & A & 18421.0295(25)    &1.59   &23.5   & $^a$P$_{0,1}$\\
3(0,3) - 2(0, 2)& E & 23679.9434(25)    &3.71   &45.9   & $^a$P$_{0,1}$\\
                & A & 23790.9359(13)    &2.33   &45.3   & $^a$P$_{0,1}$\\
 & & & & &  \\
13(*,13) - 12(*,12) & E & 90629.9722(29) &32.74 &428.0 &  \\
12(3,10) - 11(3,9) & A & 96830.8697(45) &33.60  &171.0 &  \\
14(*,14) - 13(*,13) v=1 & E & 96876.3625(94) &124.33  &463.4 &  \\
14(*,14) - 13(*,13) & E & 97354.0297(31) &37.41 &461.8 &  \\
14(*,14) - 13(*,13) & A & 97403.6841(35) &36.05 &461.6 &  \\
15(*,15) - 14(*,14) & E & 104078.1290(33) &42.40  &495.5 &  \\
15(*,15) - 14(*,14) & A & 104127.4707(38) &41.05  &495.3 &  \\
14(2,12) - 13(2,11) & E & 110443.5401(44) &45.08  &203.0 &  \\
15(2,14) - 14(2,13) & E & 110488.6550(31) &46.76  &224.7 &  \\
15(1,14) - 14(1,13) & E & 110493.3785(31) &46.76  &224.7 &  \\
16(*,16) - 15(*,15) & E & 110802.2128(35) &47.72  &529.1 &  \\
16(*,16) - 15(*,15) & A & 110851.2160(40) &46.37  &529.0 &  \\
16(*,16) - 15(*,15) v=1 & A & 111137.3603(103) &120.00  &528.2 &\\
  & & & & &  \\
18(*,17) - 17(*,16) & E & 130649.9350(34) & 64.60  & 577.4 &  \\
18(*,17) - 17(*,16) & A & 130810.0540(37) & 63.35  & 576.9 &  \\
20(*,19) - 19(*,18) & E & 144091.6978(36) & 78.11  & 644.6 &  \\
20(*,19) - 19(*,18) & A & 144251.6541(39) & 76.88  & 644.2 &  \\
21(*,21) - 20(*,20) & E & 144421.2557(45) & 79.15  & 697.5 &  \\
21(*,21) - 20(*,20) & A & 144468.4149(50) & 77.81  & 697.4 &  \\

\hline
    \end{tabular}
   \tablefoot{The approximate center frequency is listed for those transitions that are not resolved given the spectroscopic resolution of the astronomical observations. Each of these lines consists of two a-type and two b-type transitions (for example, $18_{*,17} - 17_{*,16}$ refers to the following 4 transitions: $18_{1,17} - 17_{2,16}$, $18_{2,17} - 17_{2,16}$, $18_{1,17} - 17_{1,16}$ and $18_{2,17} - 17_{1,16}$ that are nearly frequency coincident). The listed $S_{ij}\mu^{2}$ value is the sum of all four transitions in each group. This coincidence is indicated by the asterisk substituted for the K$_a$ quantum number. Uncertainties  in  parentheses refer to the least significant digit and are 2$\sigma$ values (type A coverage) \citep{JPCS_772_1_012024bib3}.}
\label{tab:thio_spec_dat}
\end{table*}

\subsection{Search for \ce{CH3CSNH2} toward NGC 6334I with ALMA \label{Sec:ngc6334i}}

To date, there are only 3 sources with confirmed detections of \ce{CH3CONH2}. These include the already discussed, Sgr B2(N) region, the high mass star forming regions NGC 6334I \citep{2020ApJ...901...37L} and G31.41+0.31 \citep{2021arXiv210711258C}. NGC 6334I, at a distance of 1.3 kpc \citep{2014ApJ...783..130R}, is a massive protocluster \citep{2006ApJ...649..888H} undergoing active star formation. Toward this source, two distinct regions exhibit spectra typical to hot core regions, designated MM1 and MM2 \citep{2012A&A...546A..87Z, 2016ApJ...832..187B}. This source was the site of the first interstellar detections of methoxymethanol (\ce{CH3OCH2OH}) \citep{2017ApJ...851L..46M} and the first two vibrationally excited torsional states of acetic acid (\ce{CH3COOH}) \citep{2019ApJ...882..118X} and has been the target of several investigations of complex molecules \citep{2019ApJ...883..129E, 2020JPCA..124..240M}.

ALMA observations of NGC 6334I at an angular resolution of 0.26\arcsec in ALMA Bands 4 and 7 were used to search for transitions of \ce{CH3CSNH2} toward this source. The details of the individual data sets and their reductions are available in \citet{2017ApJ...837L..29H, 2017ApJ...851L..46M, 2018ApJ...863L..35M}.

\begin{figure}
\centerline{\resizebox{0.98\hsize}{!}{\includegraphics[angle=0]{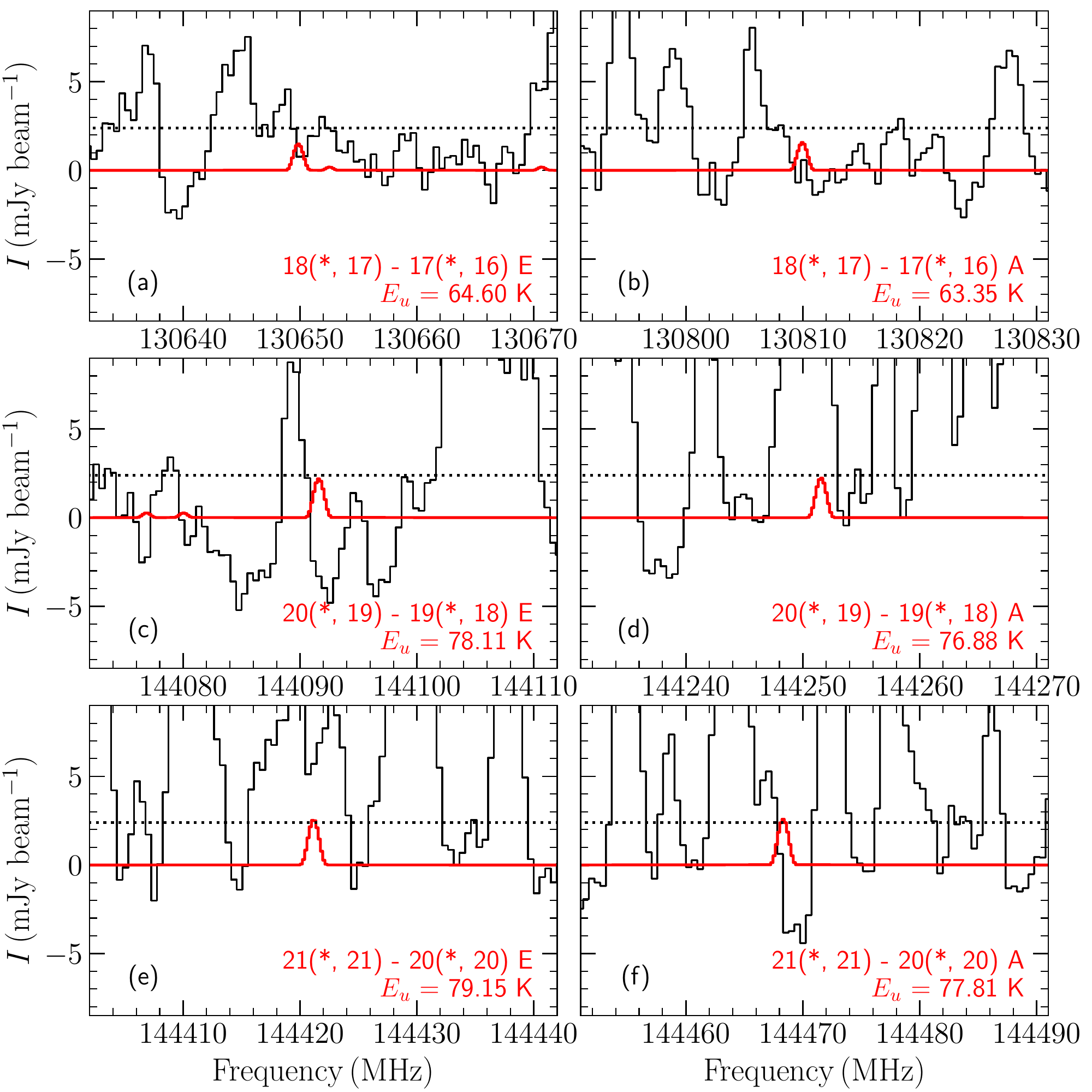}}}
\caption{Synthetic LTE spectrum of \ce{CH3CSNH2} (in red) used to derive the upper limit to its column density, overlaid on the ALMA spectrum of NGC 6334I (in black). Figure labels are similar to Figure~\ref{f:spec_ch3csnh2_ve0}. The transitions of \ce{CH3CSNH2} that are weaker than 3$\sigma$ and the ones that are heavily contaminated by other species are not shown.}
\label{f:spec_ch3csnh2_6334i}
\end{figure}

Figure~\ref{f:spec_ch3csnh2_6334i} shows the results of our attempt to detect the strongest transitions from \ce{CH3CSNH2} in ALMA Bands 4 and 7 and, in the data presented, only the most prominent transitions are shown (Table~\ref{tab:thio_spec_dat}). As with the data shown toward Sgr B2(N) with the EMoCA and PRIMOS programs, there is no evidence supporting its presence toward NGC 6334I. Similar to Figure~\ref{f:spec_ch3csnh2_ve0}, the dotted lines represent $3\sigma$ rms values and the synthetic LTE spectrum (red trace) is overlaid on the ALMA spectrum (black trace). Assuming T$_{ex}$ to be 135 K, typical of other large molecules toward this source and the best fit temperature determined from the detection of \ce{CH3CONH2} in \citet{2020ApJ...901...37L}, the upper limit of $N_T$ of \ce{CH3CSNH2} is $2\times10^{15}\ \mathrm{cm}^{-2}$. Given the measured \ce{CH3CONH2} column density of $4.7\times10^{16}\ \mathrm{cm}^{-2}$, the upper limit of the \ce{CH3CSNH2}/\ce{CH3CONH2} abundance ratio is $0.04$ - which is about a factor of 3 smaller than what was found toward Sgr B2(N2) in a similar frequency range from the EMoCA survey (See Section~\ref{Sec:sgrb2n2}).

\subsection{Cursory Search for \ce{CH3CSNH2} in the ASAI sources with the IRAM 30 m \label{Sec:asai}}

In order to complete the investigation that initially started with the search for \ce{CH3CSNH2} toward L1544 and L1157-B from the ASAI survey, the remaining targets from the ASAI survey were searched for the strongest transitions of \ce{CH3CSNH2} and \ce{CH3CONH2}.  All the objects represent different evolutionary 
stages to forming a Sun-like star - starless cores to Class 0/I sources (such as Barnard 1, NGC1333-IRAS 4A, L1157mm, L1527, and SVS13-A) as well as jet-driven shocked regions (such as L1448R2). The observational details of the ASAI survey and the sources therein are fully described in \citet{2018MNRAS.477.4792L}. 
Here we present only the parameters that were used to derive the column densities or the upper limits of both species toward these sources.

As with the searches already presented toward Sgr B2 and NGC 6334I, we find no evidence of \ce{CH3CSNH2} in any of the ASAI sources. In addition, and quite unfortunate, we also find no evidence for \ce{CH3CONH2} toward any of the ASAI sources either. As such, meaningful column density ratios cannot be determined as only upper limits to the column densities of \ce{CH3CSNH2} and \ce{CH3CONH2} can be determined.

Similar to the analysis performed toward Sgr B2(N) and NGC 6334I, a synthetic spectrum of each molecule was generated using the physical conditions for the source (Table~\ref{ASAI_params}) and the spectroscopic line parameters measured in this work. Then, following the methodology that was used to find the upper limits to \ce{HCCCHS} and \ce{HCCCHO} \citep{2020A&A...642A.206M}, the $3\sigma$ upper limit to the column density was derived, using the line that gave the most rigorous constraint (that is the line that would be the highest signal-to-noise in the event of a detection). These lines are provided in Table~\ref{ulims_table} along with the resulting upper limits.

\begin{table}[!hb]
 \begin{center}
 \caption{Parameters for each of the ASAI sources used to search for \ce{CH3CSNH2} and \ce{CH3CONH2 \label{ASAI_params}}.
 }
    \begin{tabular}{lccc}
    \hline\hline
    Source              	& $\Delta V$    		& $T_b$	        & $T_{ex}$\\
                        	& (km s$^{-1}$) 	    & (mK)                   	& (K)\\
    \midrule
    \multicolumn{4}{c}{Thioacetamide}\\
    \midrule
    Barnard 1   & 0.8   & 6.0   & 10\\
    IRAS 4A     & 5.0   & 6.2   & 21\\
    L1157B1     & 8.0   & 4.2   & 60\\
    L1157mm     & 3.0   & 7.0   & 60\\
    L1448R2     & 8.0   & 7.6   & 60\\ 
    L1527       & 0.5   & 5.6   & 12\\
    L1544       & 0.5   & 6.4   & 10\\
    SVS13A      & 3.0   & 15.8   & 80\\
    \midrule
    \multicolumn{4}{c}{Acetamide}\\
    \midrule
    Barnard 1   & 0.8   & 7.4   & 10\\
    IRAS 4A     & 5.0   & 8.3   & 21\\
    L1157B1     & 8.0   & 4.5   & 60\\
    L1157mm     & 3.0   & 7.3   & 60\\
    L1448R2     & 8.0   & 23.3   & 60 \\
    L1527       & 0.5   & 8.0   & 12\\
    L1544       & 0.5   & 8.2   & 10\\
    SVS13A      & 3.0   & 17.2   & 80\\
\hline 
 \end{tabular}
 \end{center}
\tablefoot{In all cases, a background temperature of 2.7 K was used for each source. $T_b$ is taken as the $3\sigma$ rms noise level at the location of the target line. Furthermore, all observations were taken with the IRAM 30m telescope and it was assumed that the source fills the beam except in the case of SVS13A where a source size of 0.$''$3 was used.  For additional information on each of the sources as well as a reference to the original citations where the physical conditions were measured, see \citet{2020JMoSp.37111304M}.}
\end{table}

\begin{table*}[!htb]
    \scriptsize
    \centering
    \caption{Upper limits to the abundances of \ce{CH3CSNH2} and \ce{CH3CONH2} toward the ASAI sources and the spectral line parameters used to calculate the upper limits in each of the sets of observations.}
    \begin{tabular}{l r r r r r c c c }
    \toprule
    \toprule
    Source  &\multicolumn{1}{c}{Frequency}  &Transition &$E_u$  &$S_{ij}\mu^{2}$    &\multicolumn{1}{c}{$Q$ ($Q_{rot}$, $Q_{vib}$)$^a$} &$N_T$  &$N(H_2)$   &$X_{H_2}$\\
            &\multicolumn{1}{c}{(MHz)}      &($J_{K_a,K_c}^{\prime} - J_{K_a,K_c}^{\prime\prime}$)  &(K)    &(Debye$^2$)    &   &(cm$^{-2}$) & (cm$^{-2}$)  &\\
    \midrule
        \multicolumn{9}{c}{\ce{CH3CSNH2}}\\
    \midrule
    Barnard 1   &91379.2    &$11_{2,9} - 10_{2,8}$ A    &28.9   &156.8  &807 (807, 1.00)    &$\leq 1.06\times10^{12}$    &$1.5\times10^{23}$ &$\leq 7\times10^{-12}$\\
    IRAS 4A     &104127.5   &$15_{*,15} - 14_{*,14}$ A  &41.1   &495.4  &2484 (2484, 1.00)  &$\leq 1.97\times10^{12}$    &$3.7\times10^{23}$ &$\leq 5\times10^{-12}$\\
    L1157B1     &90679.9    &$13_{*,13} - 12_{*,12}$ A  &31.4   &427.9  &16702 (16702, 1.00)&$\leq 4.61\times10^{12}$    &$1\times10^{21}$  &$\leq 5\times10^{-9}$\\
    L1157mm     &104078.1   &$15_{*,15} - 14_{*,14}$ E  &42.4   &495.6  &16702 (16702, 1.00)&$\leq 2.53\times10^{12}$    &$6\times10^{21}$   &$\leq 4\times10^{-10}$\\
    L1448R2     &97403.7    &$14_{*,14} - 13_{*,13}$ A  &36.1   &461.6  &16702 (16702, 1.00)&$\leq 7.48\times10^{12}$    &$3.5\times10^{23}$ &$\leq 2\times10^{-11}$\\
    L1527       &104127.5   &$15_{*,15} - 14_{*,14}$ A  &41.1   &495.4  &1045 (1045, 1.00)  &$\leq 4.15\times10^{11}$    &$2.8\times10^{22}$ &$\leq 1\times10^{-11}$\\
    L1544       &91492.8    &$10_{3,7} - 9_{3,6}$ A     &26.1   &142.5  &807 (807, 1.00)    &$\leq 5.49\times10^{11}$    &$5\times10^{21}$   &$\leq 1\times10^{-10}$\\
    SVS13A      &238335.5   &$34_{*,33} - 33_{*,32}$ A  &207.7  &1115.6 &29696 (29696, 1.00)&$\leq 1.82\times10^{16}$    &$3\times10^{24}$   &$\leq 6\times10^{-10}$\\
    \midrule
        \multicolumn{9}{c}{\ce{CH3CONH2}}\\
    \midrule
    Barnard 1   &87632.5    &$8_{*,8} - 7_{*,7}$ E      &19.7   &194.8  &370 (370, 1.00)    &$\leq 2.01\times10^{11}$    & $1.5\times10^{23}$    & $\leq 1\times10^{-12}$\\
    IRAS 4A     &108255.2   &$10_{*,10} - 9_{*,9}$ E    &29.7   &254.0  &1382 (1382, 1.00)  &$\leq 1.60\times10^{12}$    & $3.7\times10^{23}$    & $\leq 4\times10^{-12}$\\
    L1157B1     &108255.2   &$10_{*,10} - 9_{*,9}$ E    &29.7   &254.0  &10929 (10929, 1.00)&$\leq 4.27\times10^{12}$    & $1\times10^{21}$      & $\leq 4\times10^{-9}$\\
    L1157mm     &97943.9    &$9_{*,9} - 8_{*,8}$ E      &24.5   &226.8  &10929 (10929, 1.00)&$\leq 2.97\times10^{12}$    & $6\times10^{21}$      & $\leq 5\times10^{-10}$ \\
    L1448R2     &149411.2   &$14_{*,14} - 13_{*,13}$ A  &61.4   &361.6  &10929 (10929, 1.00)&$\leq 1.08\times10^{13}$    & $3.5\times10^{23}$    & $\leq 3\times10^{-11}$\\
    L1527       &87632.5    &$8_{*,8} - 7_{*,7}$ E      &19.7   &194.8  &500 (500, 1.00)    &$\leq 1.54\times10^{11}$    & $2.8\times10^{22}$    & $\leq 6\times10^{-12}$\\
    L1544       &97943.9    &$9_{*,9} - 8_{*,8}$ E      &24.5   &226.8  &370 (370, 1.00)    &$\leq 1.64\times10^{11}$    & $5\times10^{21}$      & $\leq 3\times10^{-11}$\\
    SVS13A      &252588.4   &$24_{*,24} - 23_{*,23}$ E  &154.4  &632.8  &19442 (19442, 1.00)&$\leq 9.79\times10^{15}$    & $3\times10^{24}$      & $\leq 3\times10^{-9}$\\
\bottomrule
\end{tabular}
     \label{ulims_table}
\tablefoot{The approximate center frequency is listed for those transitions that are not resolved given the spectroscopic resolution of the astronomical observations. Each of these lines consists of two a-type and two b-type transitions (for example, $15_{*,15} - 14_{*,14}$ refers to the following 4 transitions: $15_{0,15} - 14_{1,14}$, $15_{1,15} - 14_{1,14}$, $15_{1,15} - 14_{0,14}$ and $15_{1,15} - 14_{1,14}$ that are nearly frequency coincident). The listed $S_{ij}\mu^{2}$ value is the sum of all four transitions in each group. This coincidence is indicated by the asterisk substituted for the K$_a$ quantum number.  The total partition functions and column densities were calculated at the measured temperature of each source as listed in Table~\ref{part_funct}.}
\end{table*}

\begin{figure*}[!htb]
    \centering
    \includegraphics[width=\textwidth]{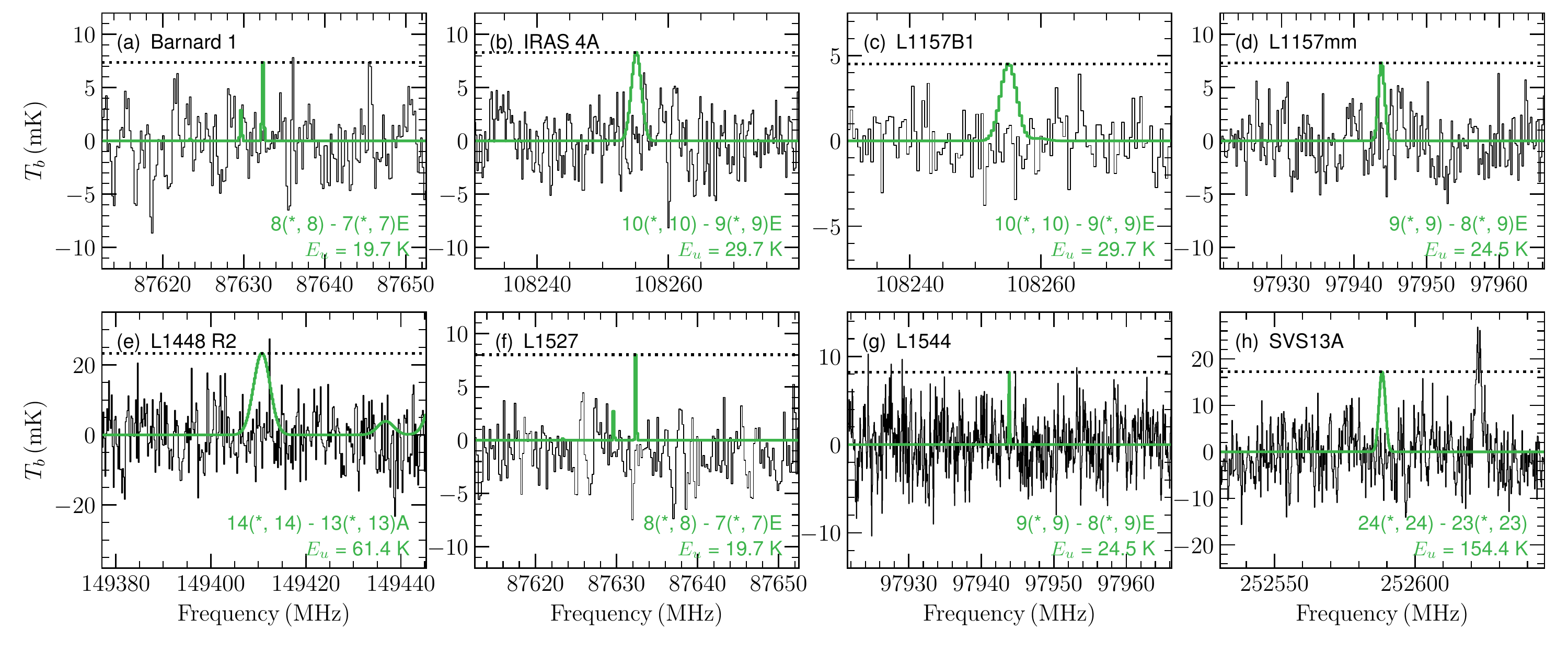}
    \caption{Transitions of \ce{CH3CONH2} used to calculate the $3\sigma$ upper limits given in Table~\ref{ulims_table}. In each panel, the red trace shows the transition simulated using the derived upper limit column density and the physical parameters assumed for that source. The dotted line in each panel indicates the $3\sigma$ noise level. The quantum numbers and upper-level energies for each transition is shown in the lower right of each panel. The source name is given in the upper left of each panel. Due to the large variances between observations, the intensity and velocity axes are not necessarily uniform between each panel.}
    \label{acet_ulim_fig}
\end{figure*}

\begin{figure*}[!htb]
    \centering
    \includegraphics[width=\textwidth]{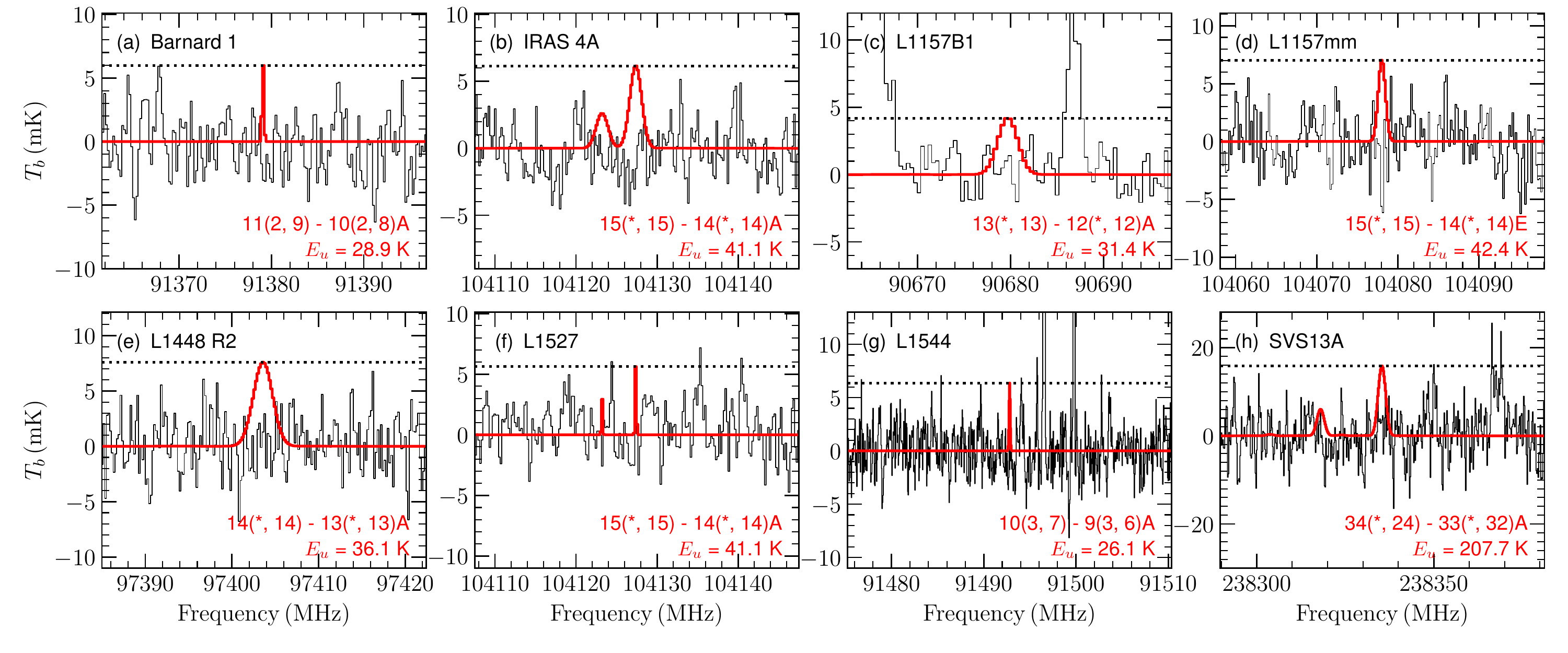}
    \caption{Transitions of \ce{CH3CSNH2} used to calculate the upper limits given in Table~\ref{ulims_table}. Panel description is the same as in Figure~\ref{acet_ulim_fig}.}
    \label{thio_ulim_fig}
\end{figure*}

\section{Discussion \label{Sec:Discussion}}

The question that remains is whether the detection (or non-detections) of these larger molecules better calibrate existing chemical formation theories.  Earlier work suggested that the formation of \ce{NH2CHO} (and possibly \ce{CH3CONH2}) can form in the gas phase via \ce{NH2} + \ce{H2CO} $\longrightarrow$ \ce{NH2CHO} + H \citep{2015MNRAS.453L..31B,2017MNRAS.468L...1S}.  An analogous reaction with \ce{CH3CHO}, that is, \ce{NH2} + \ce{CH3CHO}, can lead to \ce{CH3CONH2} + H in competition with another channel leading to \ce{NH2CHO} + \ce{CH3}. However, a kinetic study has shown that \ce{NH2} + \ce{CH3CHO} has an activation energy barrier \citep{https://doi.org/10.1002/bbpc.19860901218}.  Following these examples, the reactions \ce{NH2} + \ce{H2CS} and \ce{NH2} + \ce{CH3CHS} may lead to the formation of \ce{NH2CHS} and \ce{CH3CSNH2}, respectively. Of course, dedicated work is necessary to confirm this suggestion.

A more recent study was done to try and determine the most probable formation route to form \ce{CH3CONH2} under interstellar conditions. \citet{2018MolAs..13....1F} investigated every possible structural isomer of \ce{CH3CONH2} and as \ce{CH3CONH2} was found to be the most stable isomer, they selected the 4 most stable bimolecular species (and associated reaction pathways) as possible molecular precursors to form \ce{CH3CONH2} (as well as other possible unimolecular formation schemes whereby \ce{CH3CONH2} could form by the rearrangement of atoms from a molecular isomer - such as \ce{CH3NHCHO}). In that work, it was shown that the most probable reaction pathways to forming \ce{CH3CONH2} would be the following gas phase reactions:
\begin{alignat}{2}
    &\ce{CH4}    &&+ \ce{HNCO}\\
    &\ce{H2O}    &&+ \ce{CH3CN}\\
    &\ce{CO}     &&+ \ce{CH3NH2}\\
    &\ce{NH3}    &&+ \ce{H2CCO}. 
\end{alignat}

Of these reactions, only (3) and (5) could be feasible and the formation route via \ce{H2O + CH3CN} requires a 2-step process to form \ce{CH3CONH2} through the isomerization of \ce{CH3NHCHO}. However, both have to travel through large transition state barriers which make these routes highly improbable except in regions that can clearly drive nonthermal excitation - such as shocked regions.  For example, the formation (and detection) of methyl isocyanide (\ce{CH3NC}) and carbodiimide (\ce{HNCNH}) is explained by the nonthermal processes from the isomerization of methyl cyanide (\ce{CH3CN}) and cyanamide (\ce{NH2CN}), respectively \citep{2005ApJ...632..333R, 2012ApJ...758L..33M}.  As such, the final formation of \ce{CH3CONH2} could be through the isomerization of \ce{CH3NHCHO}.

As always, additional possibilities exist to form molecules on grain surfaces. In recent experiments by \citet{2018MNRAS.480.3628L}, a \ce{CH4}:\ce{HNCO} ice mixture at 30~K was irradiated by far UV photons to investigate the potential products generated.  The precursors used in these reactions are the same chemical precursors as in reaction (2) only now these species are contained in an ice matrix instead of in the gas phase. The main conclusion was that the formation of amides can be explained by nonenergetic radical recombination reactions with \ce{NH2}. For example, the formation of formamide (\ce{NH2CHO}) on a grain surface takes place by:
\begin{alignat}{3}
\ce{&NH2 &&+ CHO &&-> NH2CHO}
\intertext{while acetamide is formed by:}
\ce{&NH2  &&+ CO &&-> NH2CO}\\
\ce{&NH2CO &&+ CH3 &&-> CH3CONH2}.
\end{alignat}

The experiments successfully reproduced the abundance ratios \ce{CH3CONH2}/\ce{NH2CHO} measured in the ISM. The experimental value determined is 0.4$\pm^{0.39}_{0.14}$ and while the observed values vary toward different sources, all measured values fall within a range of 0.04-1.6 (for example, \citet{2006ApJ...643L..25H, 2017A&A...601A..49B, 2019A&A...628A..10B, 2011ApJ...743...60H} and this work). The conclusion reached based on these experiments was that specifically toward hot cores, the abundance of \ce{CH3CONH2} can be explained by a formation on interstellar grains with subsequent desorption from grain surfaces. Yet, gas phase reactions under some interstellar conditions cannot be ruled out. By observing less abundant species with similar molecular structure, it may be possible to better determine the formation pathways - or at least which pathway is more dominant or prevalent.

Acetamide (\ce{CH3CONH2}) and other oxygen-bearing amide molecules are shown to efficiently form on grain surfaces involving radicals. The reactant radicals as listed in reactions (6), (7), and (8) form through the photodissociation of methanol and ammonia on grain surfaces.  Given the presumed reactions (either in the gas phase or grain surface) for the formation of \ce{CH3CONH2}, it may be possible to obtain additional insight into the chemical kinetics by looking for similar reactions to form \ce{CH3CSNH2}. Yet, the overall excitation and distribution of S-bearing molecules are quite different from their O-bearing counterparts.  Gas-grain models specifically do not take into account surface reactions from many small S-bearing species as the behavior of sulfur on grain surfaces is less well understood.
Furthermore, the energetics of the S-substituted species will be different from their O-bearing counterparts making the chemical kinetics that is driving their formation much different given a similar looking reaction.

Considering that a majority of large sulfur-bearing molecules have been detected toward cold regions such as dark clouds \citep[for example,][and references therein]{2021A&A...648L...3C}, the possible gas-phase formation pathways to forming \ce{CH3CONH2} may help to guide our understanding on the formation mechanism of its sulfur-substitute amide, \ce{CH3CSNH2}. Therefore, we propose analogous neutral-neutral reactions to that of \ce{CH3CONH2}:

\begin{alignat}{2}
&\ce{CH4} &&+  \ce{HNCS}\\
&\ce{H2S} &&+  \ce{CH3CN}\\
&\ce{CS}  &&+  \ce{CH3NH2}\\
&\ce{NH3} &&+  \ce{H2CCS}.
\end{alignat}

Of the precursor species in these reactions, all have been detected in the gas phase toward a diverse sample of astronomical objects.  For example, HNCS and \ce{CH3NH2} were detected in the outer envelopes and hot cores of Sgr B2(N) \citep{1974ApJ...191L.135K, 2009ApJ...702L.124H, 2015MNRAS.452.3969C}; \ce{CH3CN} is ubiquitous toward numerous sources including hot cores \citep[and references therein]{2021A&A...648A..66G}, translucent clouds \citep{2015MNRAS.452.3969C}, and dark clouds \citep{1993MNRAS.263L..40W}. 

Finally, \ce{H2CCS} has recently been detected toward cold regions \citep{2021A&A...648L...3C} and \ce{CH4}, \ce{H2S}, \ce{CS} and \ce{NH3} are all highly abundant in the gas phase in nearly all regions of molecular gas.  As such, the proposed gas-phase formation routes to \ce{CH3CSNH2} in reactions (9-12) may be possible if energetically favorable. To test their feasibility, quantum chemical calculations were carried out with the Gaussian 16 suite of programs \citep{g16}. All geometries were optimized using Density Functional Theory with the B3LYP method \citep{1994CPL...225..247S} and a 6-311G++(d,p) basis set. The energetics were corrected with free energy corrections at a temperature of 15 K at the same level of theory. Following similar reaction paths as \citet{2018MolAs..13....1F}, transition states and intermediates were only found for reactions (10) and (12) due to the potentially large activation energies of reactions (9) and (11). The transition states were confirmed using an internal reaction coordinate analysis, and all stationary points found were further constrained by performing CCSD(T) single point energy calculations with the aug-cc-pVTZ basis set. The Gibbs free energy of reactions (9), (10), (11), and (12) were found to be -37.52, -27.62, -236.11, and -133.87 kJ/mole, respectively. Reaction (10) was found to be a two step process with Gibbs free energy of activations of 210.89 kJ/mole for the first step and 111.56 kJ/mole for the second. Reaction (12) was a single step process with a Gibbs free energy of activation of 170.62 kJ/mole. Though all proposed reactions are exothermic, the associated energy barriers are rather large for feasible reaction paths in cold interstellar conditions.

Also, the presumed main driver of S-chemistry on grain surfaces, \ce{H2S}, has yet to be unambigously detected on a grain surface.  The grain surface reactions to the formation of larger S-bearing molecules require that a radical containing S (such as, \ce{HCS}) or neutral species (such as, \ce{CS}) be available in large abundances on grain surfaces.  Since these precursor species have yet to be directly detected on grain surfaces, it is unlikely that analogous routes with S-bearing species in reactions (6-8) would lead to the formation of larger S-bearing molecules. In addition, dedicated searches for larger S-bearing molecules have taken place over a variety of sources with several non-detections (for example, \ce{CH3CHS}\citep{2020JMoSp.37111304M}, HCCCHS\citep{2020A&A...642A.206M}, and \ce{NH2CHS}\citep{2020A&A...642A.206M}) and toward locations where comparative O-chemistry may not be present \citep{2021A&A...648L...3C}. As such, it is currently difficult to constrain any of the chemistry with such a disparate set of data, yet trends comparing the relative abundances between the O-bearing vs. S-bearing species have been done toward those sources even if all we have are upper limits to the column densities which may eventually lead to a better determination of the formation pathways of both families of molecules.


\section{Conclusions \label{Sec:Conclusion}}

The spectrum of \ce{CH3CSNH2} was investigated and new fits were generated using the RAM method to assign the spectrum of \ce{CH3CSNH2} from 150 to 650 GHz.  A total of 1428 transitions from the v$_t$=0 state with maximum values J=47 and K$_a$=20  in  the  range  to  330  GHz,  and J=95 and K$_a$=20  in  the  range  400–660 GHz were assigned. We also assigned 321 transitions from the v$_t$=1 state with the maximum values J=35 and K$_a$=9 up to 330 GHz.  The final data set contains both the data from the previous study \citep{2019ECS.....3.1537M}, and new assignments in the millimeter and submillimeter-wave ranges. The total of 2035 lines for v$_t$=0 and v$_t$=1 states of A and E symmetries were fit with root-mean-square deviation of 43.4 kHz. The final fit is based on  the  RAM  Hamiltonian  model  that  includes  40  parameters.  An astronomical search for \ce{CH3CSNH2} was conducted based on all the new spectroscopic data.  No transitions of \ce{CH3CSNH2} were detected toward any source.  Using the appropriate telescope and parameters for each astronomical source, upper limits to the column densities were found for \ce{CH3CSNH2} toward each source.

In addition, new transitions of \ce{CH3CONH2} were searched for and detected toward Sgr B2(N-LMH) utilizing the GBT PRIMOS data set.  In total, the best fit model to the PRIMOS data was found for an excitation temperature of 5.8 K and a source size of $20\arcsec$ for both the A- and E-state transitions, and the total measured \ce{CH3CONH2} column density determined from the PRIMOS data from the current analysis is $\sim7.7\times 10^{13}\ \mathrm{cm}^{-2}$ for the A-state transitions and $\sim1.5\times10^{14}\ \mathrm{cm}^{-2}$ for the E-state transitions.  As expected, the observed temperature and measured column densities from this current work are much lower than what was reported by \citet{2017A&A...601A..49B} toward SgrB2(N2) from the EMoCA survey where \ce{CH3CONH2} was detected at a column density of  $\sim2\times 10^{17}\ \mathrm{cm}^{-2}$ and an excitation temperature of 180 K - clearly emission coming from a hot core region which is beam diluted in our GBT observations.  What this analysis also indicates, given the structural similarity of \ce{CH3CONH2} and \ce{CH3CSNH2}, are the types of transitions that are most likely to be detected at lower frequencies with single dish facilities such as the GBT.  And while no transitions of \ce{CH3CSNH2} were detected, the investigation of large S-bearing species compared to their O-bearing counterparts may eventually provide insight into possible formation routes.  The gas phase formation routes to form larger S-bearing molecules need further investigation into their energetics.  In addition, if grain surface reactions are to be considered as a viable route to the formation of larger S-bearing species, precursors such as \ce{H2S} and \ce{CS} need to be first identified within grain mantles.

\begin{acknowledgements}
We would like to thank an anonymous referee for their strong support of this work.  This paper makes use of the following ALMA data: ADS/JAO.ALMA\#2011.0.00017.S, ADS/JAO.ALMA\#2012.1.00012.S, ADS/JAO.ALMA\#2017.1.00370.S, and ADS/JAO.ALMA\#2017.1.00661.S. ALMA is a partnership of ESO (representing its member states), NSF (USA), and NINS (Japan), together with NRC (Canada), NSC and ASIAA (Taiwan), and KASI (Republic of Korea), in cooperation with the Republic of Chile. The Joint ALMA Observatory is operated by ESO, AUI/NRAO, and NAOJ. The National Radio Astronomy Observatory is a facility of the National Science Foundation operated under cooperative agreement by Associated Universities, Inc. The interferometric data are available in the ALMA archive at \url{https://almascience.eso.org/aq/}. This paper makes use of the PRIMOS data under GBT Archive Project Code AGBT07A-051. The Green Bank Observatory is a facility of the National Science Foundation operated under cooperative agreement by Associated Universities, Inc. This work is based on observations carried out as part of the Large Program ASAI under project number 012-12 with the IRAM 30m telescope. IRAM is supported by INSU/CNRS (France), MPG (Germany) and IGN (Spain). C.X. is a Grote Reber Fellow, and acknowledges support from the National Science Foundation through the Grote Reber Fellowship Program administered by Associated Universities, Inc./National Radio Astronomy Observatory and the Virginia Space Grant Consortium. Part of this work has been carried out within the Collaborative Research Centre 956, sub-project B3, funded by the Deutsche Forschungsgemeinschaft (DFG) -- project ID 184018867.
\end{acknowledgements}

\bibliographystyle{aa}
\bibliography{bibliography,misc,margules}

\end{document}

%% file: tab_thioacetamide_weedsmodel.tex
\begin{table*}[!ht]
 \begin{center}
 \caption{
 Parameters of our best-fit LTE model of methanol, methyl mercaptan, and acetamide, and column density upper limit for thioacetamide, toward Sgr~B2(N2).
}
 \label{t:coldens}
 \begin{tabular*}{\textwidth}{@{\extracolsep{\fill}}lcrcccccc}
 \hline\hline
 \multicolumn{1}{c}{Molecule} & \multicolumn{1}{c}{Status\tablefootmark{a}} & \multicolumn{1}{c}{$N_{\rm det}$\tablefootmark{b}} & \multicolumn{1}{c}{Size\tablefootmark{c}} & \multicolumn{1}{c}{$T_{\mathrm{rot}}$\tablefootmark{d}} & \multicolumn{1}{c}{$N$\tablefootmark{e}} & \multicolumn{1}{c}{$F_{\rm vib}$\tablefootmark{f}} & \multicolumn{1}{c}{$\Delta V$\tablefootmark{g}} & \multicolumn{1}{c}{$V_{\mathrm{off}}$\tablefootmark{h}} \\ 
  & & & \multicolumn{1}{c}{\small ($\arcsec$)} & \multicolumn{1}{c}{\small (K)} & \multicolumn{1}{c}{\small (cm$^{-2}$)} & & \multicolumn{1}{c}{\small ($\mathrm{km\ s}^{-1}$)} & \multicolumn{1}{c}{\small ($\mathrm{km\ s}^{-1}$)} \\ 
 \hline
 \ce{CH3OH}, $\varv=0$ & d & 41 &  1.4 &  160 &  4.0 (19) & 1.00 & 5.4 & $-0.5$ \\ 
 \ce{CH3SH}, $\varv=0$ & d & 12 &  1.4 &  180 &  3.4 (17) & 1.00 & 5.4 & $-0.5$ \\ 
\hline 
 \ce{CH3C(O)NH2}, $\varv=0$ & d & 10 &  0.9 &  180 &  1.4 (17) & 1.23 & 5.0 & $1.5$ \\ 
 \ce{CH3C(S)NH2}, $\varv=0$ & n & 0 &  0.9 &  180 & $<$  1.6 (16) & 1.20 & 5.0 & $1.5$ \\ 
\hline 
 \end{tabular*}
 \end{center}
 \vspace*{-2.5ex}
 \tablefoot{The parameters for methanol, methyl mercaptan, and acetamide were taken from \citet{2016A&A...587A..92M} and \citet{2017A&A...601A..49B}.
 \tablefoottext{a}{d: detection, n: non-detection.}
 \tablefoottext{b}{Number of detected lines \citep[conservative estimate, see Sect.~3 of][]{2016A&A...587A..91B}. One line of a given species may mean a group of transitions of that species that are blended together.}
 \tablefoottext{c}{Source diameter (\textit{FWHM}).}
 \tablefoottext{d}{Rotational temperature.}
 \tablefoottext{e}{Total column density of the molecule. $x$ ($y$) means $x \times 10^y$.}
 \tablefoottext{f}{Correction factor that was applied to the column density to account for the contribution of vibrationally excited states, in the cases where this contribution was not included in the partition function of the spectroscopic predictions.}
 \tablefoottext{g}{Linewidth (\textit{FWHM}).}
 \tablefoottext{h}{Velocity offset with respect to the assumed systemic velocity of Sgr~B2(N2), $V_{\mathrm{sys}} = 74$ $\mathrm{km / s}^{-1}$.}
 }
 \end{table*}